    \documentclass[fleqn,usenatbib]{mnras}
    
    \usepackage{mathptmx}
    \usepackage{amsmath}	
    
    \usepackage[T1]{fontenc}
    
    \DeclareRobustCommand{\VAN}[3]{#2}
    \let\VANthebibliography\thebibliography
    \def\thebibliography{\DeclareRobustCommand{\VAN}[3]{##3}\VANthebibliography}

    \usepackage{graphicx}	
    \usepackage{amssymb}	

    \graphicspath{{figures/}} 
    
    \newcommand{\Msun}{M$_{\odot}$ }
    \newcommand{\mum}{$\mu$m }
    \newcommand{\parcsec}{\mbox{$.\!\!\arcsec$}}
    
    \title[Benchmarking the IRDC G351.77-0.53]{Benchmarking the IRDC G351.77-0.53: Gaia DR3 distance, mass distribution, and star formation content}
    
    \author[S. D. Reyes-Reyes et al.]
    {S. D. Reyes-Reyes, $^{1,2}$
    A. M. Stutz$^{1}$,
    S. T. Megeath$^{3}$,
    Fengwei Xu$^{4}$,
    \newauthor 
     R. H. \'Alvarez-Guti\'errez$^{1}$,
     N. Sandoval-Garrido$^{1}$,
    H.-L. Liu$^{5}$
    \\
    $^{1}$Departamento de Astronomía, Universidad de Concepción, Casilla 160-C, 4030000 Concepción, Chile; reyes.r.simon.d@gmail.com\\
    $^{2}$ Max-Planck-Institute for Astronomy, Königstuhl 17, D-69117 Heidelberg, Germany\\    
    $^{3}$Ritter Astrophysical Research Center, Dept. of Physics and Astronomy, University of Toledo, Toledo, OH 43606, USA\\
    $^{4}$Kavli Institute for Astronomy and Astrophysics, Peking University, Beijing 100871, People’s Republic of China\\
    $^{5}$School of physics and astronomy, Yunnan University, Kunming, 650091, PR China\\
    }
    
    \date{Accepted XXX. Received YYY; in original form ZZZ}
    
    \pubyear{2023}
    
\begin{document}
    \label{firstpage}
    \pagerange{\pageref{firstpage}--\pageref{lastpage}}
    \maketitle
    
\begin{abstract}
     While intensively studied, it remains unclear how the star formation
(SF) in Infrared Dark Clouds (IRDCs) compares to that of nearby
clouds. We study G351.77-0.53 (henceforth G351), a cluster-forming
filamentary IRDC.  We begin by characterizing its young stellar object
(YSO) content. Based on the average parallax of likely members, we
obtain a Gaia distance of $\sim\,2.0\pm0.14$~kpc, resolving the
literature distance ambiguity.  Using our Herschel-derived N(H$_2$)
map, we measure a total gas mass of 10200~M$_{\odot}$ (within
11~pc$^2$) and the average line-mass profile of the entire filament,
which we model as $\lambda =~1660 (w/\rm pc
)^{0.62}\,\,M_{\odot}\,\rm{pc}^{-1}$.  At $w < 0.63$~pc, our $\lambda$
profile is higher and has a steeper power-law index than $\lambda$
profiles extracted in Orion~A and most of its substructures.  Based on
the YSOs inside the filament area, we estimate the SF efficiency (SFE)
and SF rate (SFR). We calculate a factor of 5 incompleteness
correction for our YSO catalog relative to Spitzer surveys of Orion~A.
The G351 SFE is $\sim 1.8$ times lower than that of Orion~A and lower
than the median value for local clouds. We measure SFR and gas masses
to estimate the efficiency per free-fall time, $\epsilon _{\rm ff}$.
We find that $\epsilon_{\rm ff}$ is $\sim$1.1 dex below the previously
proposed mean local relation, and $\sim\,4.7\times$ below Orion~A. These observations indicate that local SF-relations do not capture variations present in the Galaxy.  We speculate that cloud youth and/or magnetic fields might account for the G351 inefficiency.
    \end{abstract}
    
    
    
    \section{Introduction} \label{section:intro}
    
    The identification of mid-infrared dark clouds (IRDC) seen in silhouette against the stars and nebulosity of the galactic plane revealed a population of clouds that  potentially contain the earliest stages of cluster and high mass star formation (M$\, > 8 $~M$_\odot$) \citep[e.g.][]{egan1998,carey1998,Rathborne2007,Sanhueza2012,Sanhueza2019,Li2023}. Although subsequent studies showed the sample of IRDCs to be heterogeneous and dominated by clouds that will not form high mass stars \citep{kauffmann10,Morii2023}, massive IRDCs may form in some cases large clusters of low and high mass stars, as found in the Orion~A cloud \citep{Motte2018, stutz16, Megeath_2022}. Existing surveys in the visible, IR, and submillimiter (sub-mm), now give us the tools needed to compare IRDCs to nearby, star forming, molecular clouds. Via such comparisons, we can better establish the differences between IRDCs and nearby clouds, and establish whether IRDCs are similar to clouds in the nearest 1.5~kpc, or whether they are deficient in star formation.  If they are indeed deficient, this would imply they are either in the early stages of star formation or that they have systematically different star formation properties, presumably due to environmental factors (e.g. magnetic fields, turbulence).   
    
    Given that low to intermediate mass stars dominate the number of stars formed in molecular clouds, and given that they $-$ in contrast to high mass stars $-$ form in a wide range of gas environments, low to intermediate mass YSOs provide an excellent tracer of star formation \citep{Megeath_2022}. Unlike high mass star formation, which is often accompanied by bright extended mid-IR nebulosity \citep{Motte2018,povich13}, low to intermediate mass YSOs are best identified in point source surveys.  Wide field mid-IR imaging with Spitzer and the Wide-field Infrared Survey Explorer (WISE; \citealt{wright10}) provide the means to reliably identify low to intermediate mass YSOs in IRDCs.  In particular, the Galactic Legacy Infrared Midplane Survey Extraordinaire (Glimpse) survey \citep{benjamin2003,churchwell2009} has provided 2$\arcsec$ angular resolution image of the plane of our galaxy.  Built on these surveys, the SPICY catalog uses a machine learning approach to a combination of Glimpse and near-IR surveys to identify protostars \citep{kuhn21spicy}. For an all sky view, \citet{marton19} combined the AllWISE \citep{cutri13} and GAIA \citep{gaia2016,gaia2022} catalogs to identify YSOs, although with a significantly lower angular resolution (6$\arcsec$) and sensitivity compared to Spitzer.  From the number of YSOs detected in these surveys,  star formation rates (SFR) can be determined for the IRDCs. These can then be compared to the SFRs of nearby clouds without relying on comparatively sparse and poorly characterized high mass stellar populations.  
    
    Turning to the gas mass reservoir, which must be characterized in star and cluster forming systems, far-IR and (sub-)millimeter surveys complement these SFRs by mapping the structure of the cold, molecular gas over molecular cloud scales. In particular, the ATLASGAL survey, the Planck Survey, and Herschel Infrared Galactic Plane survey \citep[Hi-GAL,][]{Molinari2010,Molinari2016} data can be used to make maps of the structure of molecular clouds through their thermal dust emission \citep[e.g.,][]{laun13,Stutz_2013,stutz15,csengeri16, alvarez}.  These can be compared to maps of nearby clouds using various methods for characterizing filamentary clouds \citep{stutz15,stutz16}. More fundamentally, any mass measurements directly depend on the distance estimation to these clouds, and here the accurate astrometry of the Gaia survey have been playing a major role during the last decade by providing the means for establishing distances to regions if their embedded members can be found in the Gaia archive. 
    
    In combination, the distances, SFRs, and gas column densities can be used to directly compare star formation in IRDCs to that in nearby clouds \citep{pokhrel21,Megeath_2022}. Recently, surveys of YSOs of nearby molecular clouds have found a significant degree of uniformity in local clouds \citep{Gutermuth_2011,  Lada_2013, pokhrel20, pokhrel21}.  In particular, \citet{pokhrel21} establish a relation where the star formation rate efficiency per local free-fall time ($\epsilon_{\rm ff}$) is 2.5\% (also see \citealt{hu21}), which is established in a sample of clouds ranging from low mass star forming clouds at 140~pc to high mass star forming clouds at 1.4~kpc. Although still under debate \citep{Lada_2013,evans14,Ginsburg2018}, these star formation relations (SF-relations) provide a benchmark for comparing the SFRs in IRDCs to those in nearby clouds. 
    
    G351.77-0.53 (henceforth G351) is a filamentary IRDC that has been studied through a series of primarily millimeter observations. Despite the attention, the large uncertainty in the distance, with estimates ranging from 1-2~kpc, results in significant uncertainties in the inferred properties. 
    Previous estimations of the G351 filament are focused on the protocluster area and are based on Galactic rotation curve models. \cite{motte22} reported a distance $D=2$~kpc from the BeSSeL kinematic distance calculator, which is an average between the two equally probable solutions 1.3 and 2.7~kpc retrieved by the calculator. \cite{leurini11a} reported a distance  $D \le 1$~kpc using velocities based on thermal lines. Other distances $D=$~0.7,~1,~2.2~kpc are also reported in the literature based on maser emission lines (\citealt{forster-caswell,macleod98,norris93,miettinen06}). 
    
    \begin{figure*}
    	\includegraphics[width=\textwidth]{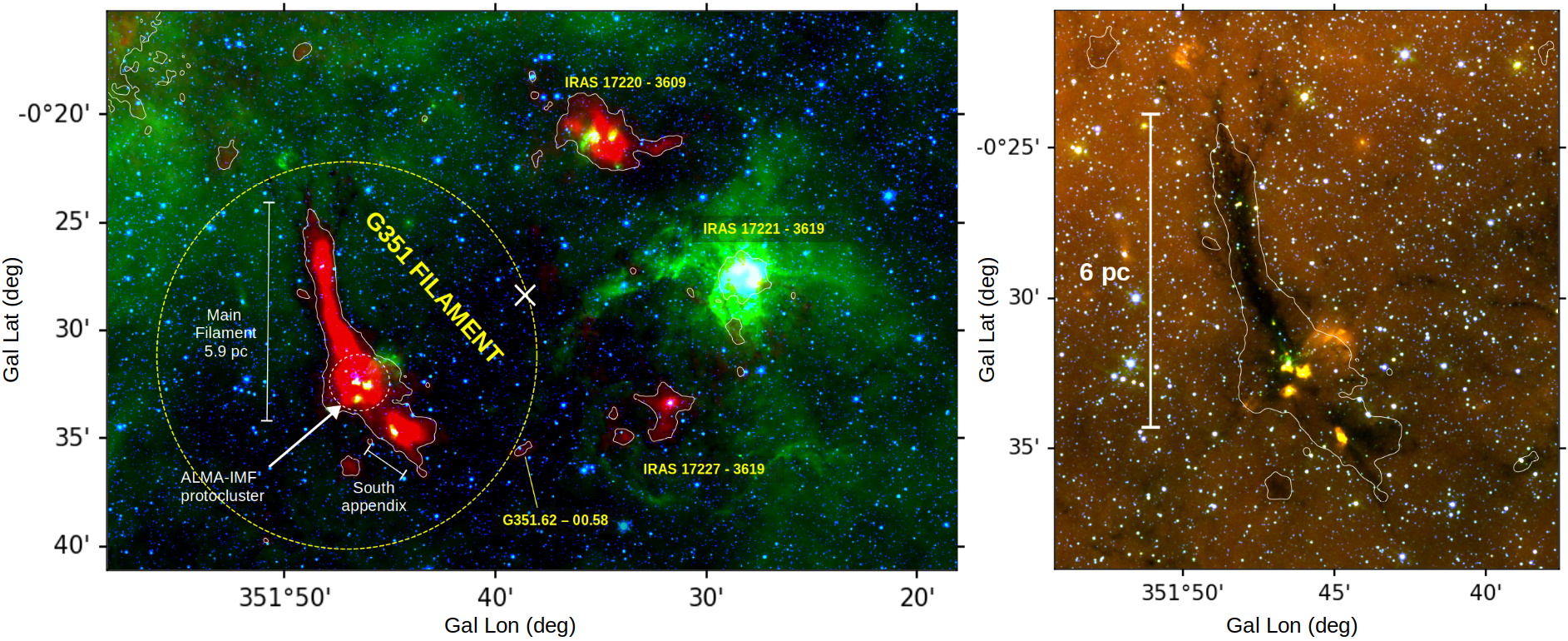}
        \caption{Left: Overview of the G351 Environment in a composite color map. The APEX+Planck dust emission at 870~\mum is shown in red, the IRAC 8.0~\mum in green, and the IRAC 3.6~\mum in blue. An APEX+Planck contour level of 0.75 Jy/beam is drawn in white, and notice that it also covers the source IRAS 17221$-$3619, revealing that the dust emission is also bright in that area. The white x-symbol shows the center of our catalog searches for the G351 Environment, and the larger circle shows the area used to retrieve catalogs for the G351 Filament. Right: Zoom-in to the G351 Filament field in a color composite of three IRAC bands, where 8.0, 5.8 and 3.6~\mum are shown in red, green and blue, respectively. The contour represents the same emission as in the left panel. This panel highlights, in an arbitrary color scale, that the cloud appears dark in the mid-infrared. The bright nebulous region locates the active protocluster, where cluster and massive star formation is ongoing.}
        \label{fig:overview}
    \end{figure*}
    
    We study two fields in the G351 IRDC. Firstly, a larger scale field that captures the filamentary body in which the protocluster is embedded and the curved structure of small clouds that follow the filament from south to west. All together form a non-continuous semi-circular path in the sky at 870~$\mu$m that plausibly belongs to the same molecular environment (but see below). We will call this entire field the "G351 Environment".  Secondly, we define a small sub-field of the G351 Environment that is centered on the G351 protocluster. This includes the filamentary body in which the protocluster is embedded, together with the small vestigial appendix seen just below. This field will be called the "G351 Filament"  hereafter and it is the structure that this work studies in detail. 
    Given the intense interest in protocluster formation specifically, the ALMA Large Program ALMA-IMF \citep[e.g.][Sandoval-Garrido et al., in prep.]{motte22,ginsburg22, cunningham23} focuses on the G351 protocluster itself (along with 14 other Milky Way protoclusters) at high resolution at 1~mm and 3~mm.  Hence, the G351 filament itself can be subdivided into different components. These are the filamentary body stretching northward from and including the protocluster, which we will call the Main Filament (MF), and the small vestigial appendix seen immediately south of the protocluster, which we will call South Appendix (SA). See Fig.~\ref{fig:overview}  for an overview of the G351 Environment and G351 Filament.
    
    Line velocity measurements across the entire G351 Environment suggest that most of the bright emission at 870~\mum are associated. \cite{leurini11b} spectra of  $^{13}$CO (2$-$1), C$^{17}$O (2$-$1) and C$^{18}$O (2$-$1) reveals that the two velocity components $\sim -3$~km\,s$^{-1}$ and $\sim -22$~km\,s$^{-1}$ are common for three bright sources in this path (SA, G351.62 and IRAS 17227$-$3619), while the velocity $\sim -3$~km\,s$^{-1}$ is detected across all the MF and the velocity $\sim -22$~km\,s$^{-1}$ is detected in IRAS 17221$-$3619.  Furthermore, \cite{ryabukhina2021} spectra of CO (2$-$1) completes this picture by detecting the same two velocities in one clump of the MF and the SA (clumps 6 and 5 in figure 5 of \citealt{ryabukhina2021}). 
    
    Here, we undertake a systematic investigation of the G351 cloud and its sub-regions (see above) to compare both the gas mass reservoir and star formation properties with nearby clouds, such as Orion~A and California.  We selected as our principal reference star forming site the Integral Shaped Filament (ISF) in the Orion~A cloud for two reasons. It has a filamentary structure of similar length ($L_{ISF}=$~7.3~pc; \citealt{stutz16}) as G351, and it is a well-studied high mass star forming site, both in terms of the global YSO content and the gas properties \citep[e.g.,][]{megeath12,megeath_2016,stutz16,stutz18,sadavoy16,kainul17,gonzalez-lobos}. Hence, the ISF is currently the prime candidate to establish comparisons regarding star and cluster formation in filaments.  We also compare to the California L1482 cloud \citep{alvarez} because it has a similar mass to that of Orion~A but lower star formation activity and hence might span a lower or early phase toward cluster formation \citep{lada09}.  
    
    This paper is organized as follows.  We begin by determining the Gaia-based distance to G351 by applying techniques for identifying members in the Gaia DR3 catalog.  We then determine the cloud gas properties using data from the ATLASGAL, Planck, and Herschel surveys. We focus on the gas line-mass {\it profile} of this filamentary cloud, which can be directly compared to that of the Orion~A and California clouds. We then show that at our established distance, G351 exceeds the line mass of the ISF, the most active star forming region within 1~kpc of the Sun \citep{megeath_2016}.  Finally, we show that the apparent star formation, averaged over the $\sim\,$2.5~Myr average lifetime of dusty YSOs, is deficient compared to the Orion~A cloud, even with completeness corrections.  We discuss observational biases that could result in this underestimate, and the implications for a difference in the intracloud SF-relation. 
    
    \section{Data} \label{section:data} 
    
    Several archival data sets were used for this work, including point source catalogs built upon visible and infrared data, and imaging data in the infrared and sub-millimeter. We overview these data sets in this section.
    
    \subsection{Point source catalogs}
    
    Point sources were retrieved using cone searches as follows: the G351 Filament field was centered at the coordinate 
    RA~=~$7:26:42.620$, DEC~=~$-36:09:22.473$,
    with a radius of 0.15~degrees (dashed circle in Fig.~\ref{fig:overview}), and the G351 Environment field was centered at the coordinate RA~=~$17:26:04.9607$, DEC~=~$-36:13:39.055$ with a radius of 0.22 degrees (the $\times$-symbol in Fig.~\ref{fig:overview} shows the center of the search area). 
    
    We use the Gaia Data Release 3 (Gaia DR3) to leverage its accurate parallaxes $\varpi$ and proper motions~$\mu$. We also use the Glimpse catalog because it provides high sensitivity infrared photometry. Specifically, we use the Glimpse II Epoch 1 December '08 Archive version because it provides the largest number of sources over the G351 region.  
    
    \subsubsection{YSO Catalogs} \label{sub:YSO-catalogs}
    
    We use the three public catalogs that have identified YSOs in the G351 Filament and Environment. They are i) the \citet{kuhn21spicy} catalog, which identified 75 YSOs in the G351 Filament using a machine learning approach applied to the Spitzer/IRAC photometry combined with 2MASS and other ground-based near-IR photometry, ii) the \citet{chen2013} catalog of extended green objects (EGOs), which identified 4 objects (all along the higher N(H$_{2})$ density filament) also using Spitzer imaging from the Glimpse survey, and iii) the \citet{marton19} catalog, which identified several YSOs in a probabilistic manner.
    In particular, the probabilistic nature of the latter motivates us to briefly describe its more relevant parameters as we will use them to filter its data. \citet{marton19} identified YSOs based on Gaia DR2 crossmatched with AllWISE. They first determined the probability for the 12 and 22~$\mu$m AllWISE bands to be real detections (P$_{R}$). Then, they defined two types of YSO probabilities: one based on all four WISE bands (PL$_{Y}$), provided P$_{R} >0.5$; and the other based on the two shorter wavelength bands (PS$_{Y}$), provided P$_{R} \le$~0.5. 
    
    \begin{figure}
    	\includegraphics[width=\columnwidth]{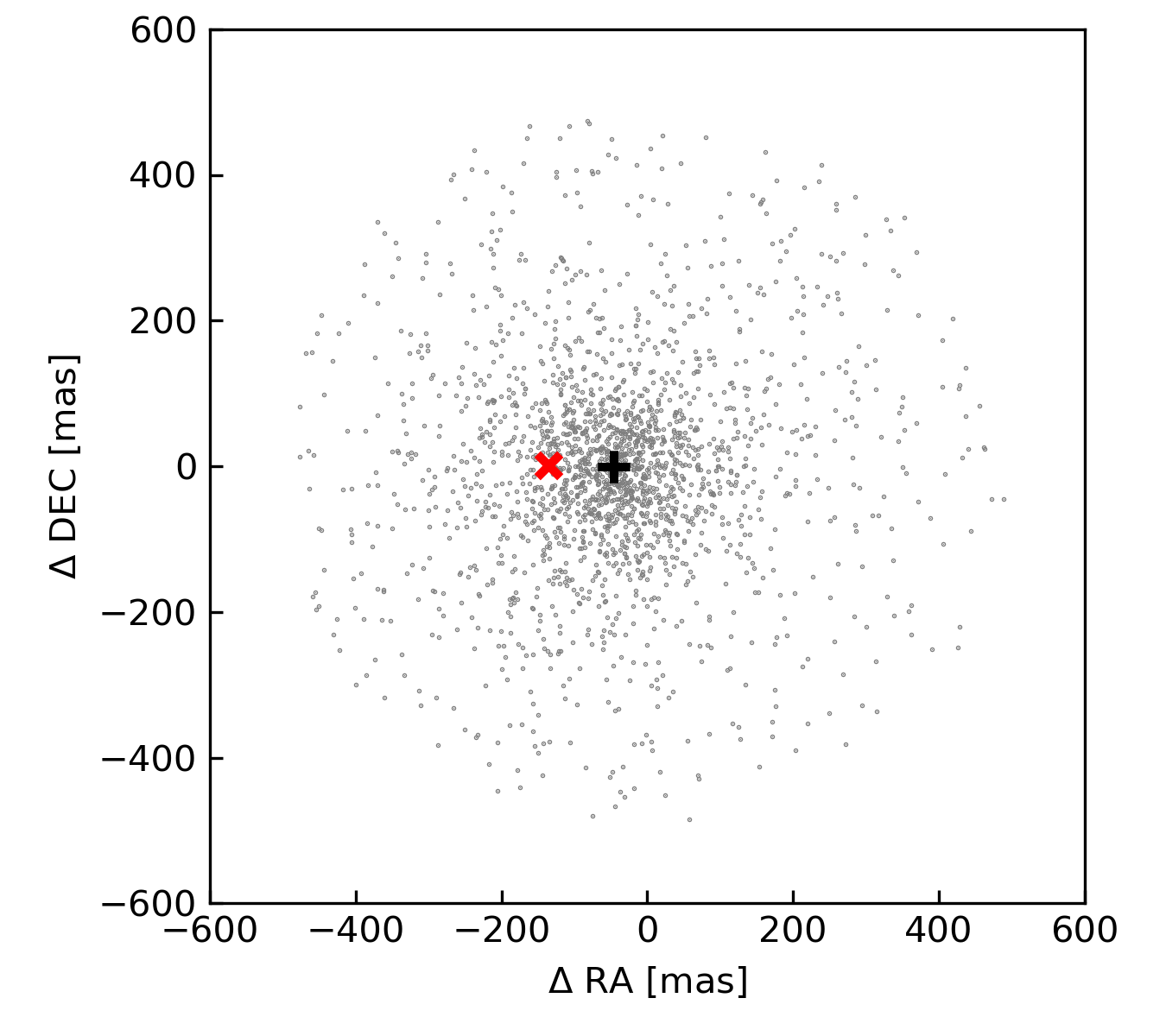}
        \caption{Example of our 2-step crossmatch process for Gaia DR3 and Glimpse in the area of the G351 Filament. We show the separations of the matched sources after correcting the relative positional shift  $\left< \Delta \rm{RA} \right>$ and $\left< \Delta \rm{DEC} \right>$ between the two catalogs, where the red $\times$-symbol represents the original positional shift $\left< \Delta \rm{RA} \right>= -135$~mas ; $\left< \Delta \rm{DEC} \right>= 1.04$~mas, and the black plus-symbol shows the subsequently reduced shift $\left< \Delta \rm{RA} \right>= -45$~mas; $\left< \Delta \rm{DEC} \right>= -1.23$~mas.}
        \label{fig:crossmatch}
    \end{figure}
    
    \subsubsection{YSO identification}\label{method:YSOidentification}
    
    In addition to these three catalogs, we independently identify YSOs following the phase-1 method (ph1) described in \cite{gutermuth09} and in \cite{gutermuth10_erratum}. This method  identifies YSOs using several criteria that use the four band 3.6-8~\mum photometry. As a first step, it identifies contaminants such as galaxies with bright emission in the PAH features, knots of shock excited emission, PAH-contaminated apertures, or active galactic nuclei (AGNs). Then, from the remaining sources, it identifies Class I or Class II YSOs. We apply these criteria to sources with photometric uncertainties $\sigma < 0.2$~mag only. 
    In the G351 Filament field we identify 61 YSOs, and after merging these YSOs to the existing YSO catalogs described above, we obtained 106 YSOs (used in Section~\ref{method: SFR-SFE}). In parallel, for the G351 Environment the final merged catalog has 225 YSO candidates (used in Section \ref{distance:method2}). 
    
    \subsubsection{YSO Catalog for Orion A} 
    \label{subsub:Megeath12}
    
    We use the \citet{megeath12,megeath_2016} YSO catalog. It provides near- and mid- infrared photometry for Orion A and B sources, classifying them as e.g. protostars or disks. It is the most complete YSO catalog over this area  and we particularly use their \textit{reliable} protostars in Section \ref{subsection:SFR}.
    
    \subsection{Imaging}
    We use the APEX+Planck dust emission map that results from combining Planck/HFI and APEX/LABOCA data from the ATLASGAL survey (\citealt{csengeri16}). This data set is publicly available\footnote{https://atlasgal.mpifr-bonn.mpg.de/cgi-bin/ATLASGAL\_DATABASE.cgi}. 
    We also use Herschel images at 70, 160, 250, 350 and 500~\mum from the Hi-GAL survey. 
    Lastly, we use images at 3.6, 5.8, and 8~\mum from the IRAC camera on board of the Spitzer telescope.
    
    \section{Gaia distance to G351} \label{section:distance}
    
    The distance to the G351 protocluster is poorly constrained in the literature with differences of factor of $\sim$2 in the reported  values (see above). We estimate a distance based on the astrometry of Gaia DR3 assuming that a molecular cloud hosting several stars will exhibit clumping of sources in proper motion and parallax space. In this regard, we adopt two approaches: 
    In method 1, we use the astrometry of Gaia sources detected in the area of the G351 Filament. In method 2, we use the astrometry of previously identified YSOs but for the area of the G351 Environment. 
    On one hand, method 1 scrutinizes the exact area of the G351 cloud, but many foreground and background sources are included.
    On the other hand, method 2 relies on objects that are much more likely to belong to the cloud and its  environment, but at the same time we lack evidence that the G351 Environment actually encloses physically associated clouds (see the above discussion).
    Despite these differences, the two methods retrieved similar results compared to the large spread in the literature. The method 1 and 2 distances are consistent within their uncertainties, but we note that the method 1 distance is $\sim$15\% larger. 
    In the context of a large disagreement in the literature, this difference is small. We adopt the average between our two results (2.109 and 1.835 kpc, see details below), and report a final constrained distance $D\simeq2$~kpc with a rounded dispersion of $\pm 0.14$~kpc for the G351 protocluster and its filament. 
    
In both methods we calculated distances using the \textit{Kalkayotl} code \citep{Olivares2020}, which is created to estimate distances to (and sizes of) clusters,  using \textit{family priors} in a bayesian analysis. This method is expected to retrieve reliable distances within $<5$~kpc (and reliable cluster sizes within $\leq 1$~kpc). We applied the most simplistic "Gaussian" family prior, where we set its median to 1.517~kpc, and its standard deviation to 0.628~kpc. The median here represents the mean value of distances reported in literature, while the standard deviation encloses all those values within two sigma  (see Section~\ref{section:intro}).  This is indeed a wide gaussian distribution, hence this bayesian result might not differentiate much from a distance obtained using the inverse of the parallax mean if the parallax dispersion of the sample is much lower (as in Method 1). The \textit{parametrization} was set to "non-central".

    We selected a strict threshold of 0.5$\arcsec$ for all the catalog-crossmatches we performed in this work. In addition, for each crossmatch we applied a correction consisting in two steps. First, by running the crossmatch the first time we obtained the systematic positional shift between the catalogs, which is revealed by the average separation of the matching sources for each coordinate, i.e. $\left< \Delta \rm{RA} \right>$ and $\left< \Delta \rm{DEC} \right>$. Then, we subtracted both quantities to one of the catalogs and reran the crossmatch. As a result, the updated mean positional shifts get reduced (sometimes significantly) and, in some cases, the number of matched sources increased. This way our crossmatches account for positional shifts that are sometimes present between some catalogs. 
    Figure \ref{fig:crossmatch} shows this process for the case with the most prominent shift, the crossmatch between the Gaia DR3 and Glimpse samples over the G351 Filament (see Method~1 below). Both catalogs have a systematic positional shift relative to each other, and that average shift is shown by the red x-symbol. At this point before correction, the number of matched sources was 1975. After correction, new matches were found, reaching a total of 2024. The updated average shift has a significant reduction as shown by the plus-symbol, although it does not reach the origin exactly $-$ as it may be expected $-$ because of the new matches that are found. 
    We also corrected all parallaxes  in our data from a systematic bias following the recipe of \cite{lindegren21}. They found that this bias produces an offset of a few tens of milliarcseconds, for which the main dependencies are the magnitude, colour, and ecliptic latitude. We call these newly corrected parallaxes as $\varpi'$ hereafter. Their recipe for correction is publicly available\footnote{The \textit{zero\_point} python package. Example code in ESA website: https://www.cosmos.esa.int/web/gaia/edr3-code}.
    
    A caveat to consider on both distance methods is a possible selection effect due to the Gaia limitations to detect the most redder sources, which might bias our results towards a closer distance. YSOs are inherently red and, combined with a high typical extinction reaching $A_V \sim 100$~mag in dense cloud regions (estimated from our N(H$_2$) map in Section \ref{method:NH2}, and figure 1 in \citealt{Draine2009}), they should give an optical telescope like Gaia preferential access to the closest and less embedded objects. We note, however, that our strict catalog crossmatches provide the certainty that the astrometry we are using actually corresponds to the infrared detected sources we use in method 2 (and also in our alternative approach in method 1).

    \begin{figure}
    	\includegraphics[width=.98\columnwidth]{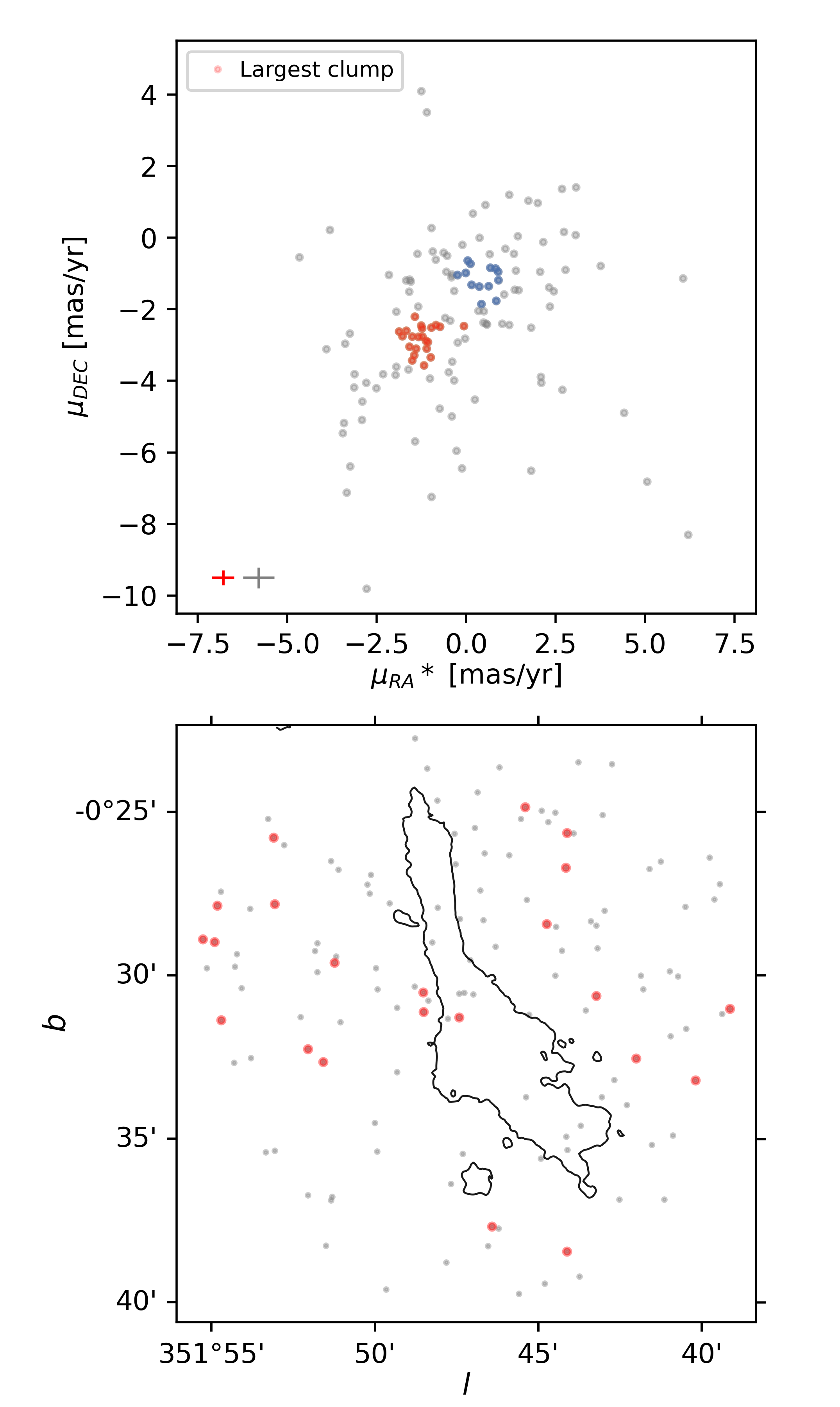}
        \caption{Gaia DR3 sources with parallax  $\varpi '$ between 0.45 and 0.50~mas. Top: The red and blue highlighted groups are those identified by HDBSCAN to be clumped in proper motion and parallax space. The red group is the largest one and we use it to estimate a distance.
        The bottom left error bars show the mean proper motion error of the largest clump (red), and the entire $\varpi'$ slice (grey). Bottom: Galactic positions of this $\varpi '$ slice sources, where we also highlight those of the largest clump (in red). The G351 filamentary structure is represented as an APEX+Planck contour level of 0.75 Jy/beam as in Fig.~\ref{fig:overview}. }
        \label{fig:mosaic1}
    \end{figure}

\subsection{Distance method 1} \label{distance:method1}
    
        We start with the Gaia DR3 raw catalog over the G351 Filament field, which contains 3831 sources (see Fig.~\ref{fig:overview} for an overview of this field and Section \ref{section:data} for the parameters of the searched area). 
    We adopt the threshold $RUWE < 1.4$ to keep those sources with well-behaved astrometrical solutions  \citep{Gaia_Technical_Note}, 
     producing an astrometrically-reliable dataset of 3178 sources that we use to identify clumped sources in proper motions and parallax.    
     To formally identify clustering we use HDBSCAN \citep{campello_2013}, a hierarchical and density-based clustering algorithm capable of dealing with noise, and  where only one single parameter (\textit{min\_cluster\_size}) is left for the user determination, making it more straightforward than other implementations like e.g. DBSCAN that uses two user-free parameters. Additionally, the probed higher sensitivity of HDBSCAN for finding clustering (although with higher false-positive detection rates; \citealt{hunt21}) makes it more suitable for detecting small groups of sources embedded in noisy data (field stars).

     We sliced this large Gaia DR3 sample in $\varpi'$ ranges of 0.05~mas to analyze individually their distribution in proper motion space and search for any possible clustering. The reason to use several slices instead of the entire dataset at once is that in the latter case HDBSCAN does not successfully identify any clustering in the data due to the high noise of the field sources, i.e. only a small fraction of sources are in a cluster, then the algorithm  cannot distinguish them as a significant overdensity over the densest region of the entire dataset, and sometimes even reporting the densest region itself as a cluster. This behavior of HDBSCAN is already reported in literature \citep{hunt21} and must be always taken into account. The slices in parallax ranges are then the best way to clean the noise around any small cluster, but also to allow the human eye to participate in their identification.

By inspecting each parallax slice, we observe only one of them  exhibiting a clear cluster. 
This particular slice $\varpi ' \in $ [0.45,0.50]~mas contains 133 sources, and we specifically identify its overdensity using the python implementation of HDBSCAN \citep{McInnes2017}. We provided our dataset of reliable astrometry as a 3D matrix input ($\varpi,\, \mu_{\rm RA},\, \mu_{\rm DEC} $), and we set the higher possible $min\_cluster\_size$ value for which the algorithm finds clustering (i.e. $min\_cluster\_size=12$). 
We expect the latter reduces false positives in each detected cluster (although slightly compromising the number size).
In general, $min\_cluster\_size$ values close to 10 can extract clusters with better sensitivity than higher $min\_cluster\_size$ values like 20 or 80 \citep{hunt21}, however, this might also bring many false positives. 

As a result, two groups are highlighted as clusters in this parallax slice. The largest one contains 22 sources (red in Fig.~\ref{fig:mosaic1}), and has a \textit{mean\_member\_probability} = 0.93. Because this is a hard cut of the data, we also explored slightly shifted slices to analyze the persistence of this particular cluster. We shifted slices of equal width (0.05~mas) in steps of 0.01~mas. Then, using the same set of initial parameters, we observed that the cluster quickly loses significance the farther a slice moves from the original range $\varpi ' \in $ [0.45,0.50]~mas. It follows that this original slice exhibits the most prominent cluster, and that the cluster cannot be found elsewhere.
Conversely, the smaller cluster contains 13 sources (blue in Fig.~\ref{fig:mosaic1}) but it is easily confused with noise when observed by the human eye, hence we discard it from this analysis (this also highlights the importance of assessing these algorithm's results together with our human pattern-recognition capabilities).
Consequently, we use the 22 members of the largest cluster to estimate a distance (refer to Table \ref{table:sources_method1} in the appendix for the astrometry used for this sample). Using \textit{Kalkayotl}, we obtain a distance estimation $D= 2.109^{+0.267}_{-0.220}$~kpc.

This result is not significantly different than the $\simeq 2.12$~kpc that we obtain by directly adopting the inverse of the error-weighted mean parallax $1/\langle \varpi ' \rangle $. Going further with this sample, the bayesian geometric distance of the \citealt{Bailer-Jones_2021} catalog suggests $D\simeq2.54$~kpc (we do not explore their "photogeometric" distance because it assumes no reddening in the line of sight, while G351 is highly extincted).
We finally note that we did not employ signal-to-noise cuts in parallax because they produce a bias towards closer distances, where uncertainties are smaller.

    The bottom panel of Fig.~\ref{fig:mosaic1} shows the positions of the clustered sources of the top panel. Despite sharing very similar proper motions, they do not appear spatially clustered or tightened to the filament. It should not be expected, however, that these sources appear highly clumped in position as they do in proper motion, because the cloud is dark for telescopes in the optical ($A_V \sim 100$), as shown by Fig.~\ref{fig:overview} (right panel), and a degree of spatial dispersion around the cloud would result from a somewhat more evolved population of stars associated with the G351 filament. Finally, we note that a possible second clump appears in proper motion space for the slice $\varpi' \in [0.55,0.60]$~mas suggesting a distance $D\simeq 1.61$~kpc, however, it is not as evident as the one we report in this work.

    In an alternative approach to this method we introduced Spitzer point sources by crossmatching our Gaia DR3 sample with Glimpse. This intends to filter out from the Gaia DR3 sample those objects that are bright in the visual but faint in the infrared, and therefore $-$~conveniently~$-$ raising the proportion of reddened sources. In that case, we would again observe that the slice $\varpi ' \in $ [0.45,0.50]~mas exhibits the cleanest clump, but this time it is less dense because it contains only 16 of the 22 sources we initially obtain without crossmatching Gaia to Glimpse. We obtain then a distance $D\simeq2.13$~kpc, slightly higher than the value we report above. In summary, this alternative procedure for method 1 introduces one extra step that barely modifies the distance estimation, but we believe it is reasonable to discuss. 

    \begin{figure}
    	\includegraphics[width=\columnwidth]{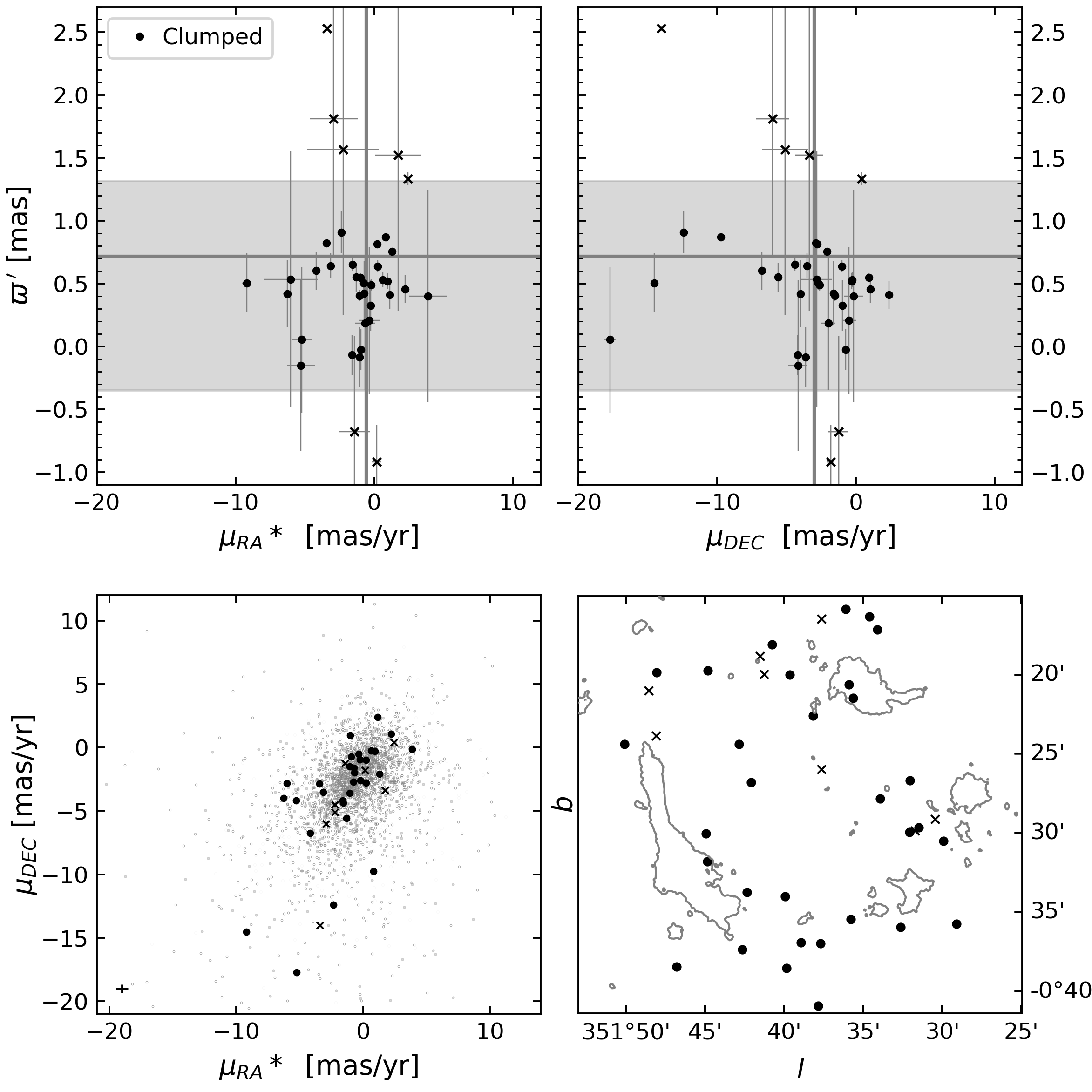}
        \caption{Astrometry of the Gaia DR3 detected YSOs in the G351 Environment region. 
        Top panels: corrected parallaxes versus Right Ascension (left) and Declination (right) proper motions. The shaded area shows the interquartile range in parallax whithin which sources are used to estimate a distance. They are shown as solid dots, while those outside are considered outliers and are shown as x-symbols. We maintain this symbology in the bottom panels. The horizontal and vertical lines crossing both frames show the mean $\varpi '$ and mean $\mu$, respectively, of the clumped sources (i.e. the solid dots) only for reference. The grey bars are the 1-$\sigma$ uncertainties in $\varpi '$ and $\mu$ of each YSO.
        Bottom left: proper motions against the whole Gaia DR3 raw catalog (grey background). The small bars at the bottom show the mean  $\sigma_{\mu_{RA}}$ and $\sigma_{\mu_{DEC}}$ of the clumped YSOs. 
        Bottom right: galactic positions of the YSOs. Contours represent an APEX+Planck emission level of 0.75 Jy/beam as in Fig.~\ref{fig:overview}.}
        \label{fig:mosaic_kuhn}
    \end{figure}
    
    \subsection{Distance method 2} \label{distance:method2}

    Here we consider only YSOs with IR excesses from dusty disks or envelopes to increase the certainty that the sources under study are associated with the cloud, following the strategy of \cite{kuhn21}. Their approach has proven to work efficiently in Galactic clusters with parallax reaching $\varpi \sim 0.5$~mas on average (Kuhn 2022, private communication). As the number of YSOs is small over the G351 Filament, we expanded our search field to include the few clouds that are plausibly associated to the same molecular environment as that of the filament (i.e. the G351 Environment field). 
    
    This larger-field YSO sample has 243 YSOs (Sections \ref{sub:YSO-catalogs} and \ref{method:YSOidentification}), but it gets significantly reduced to 47 sources when crossmatched with Gaia DR3 because of the high extinction in the denser regions ($A_V \sim 100$). For this we kept only those \citet{marton19} YSOs whose probability to be actual YSO was larger than 0.7. This means that PL$_Y \ge~0.7$ when P$_{R}>$ 0.5, and PS$_Y \ge 0.7$ when P$_{R} \le$ 0.5 (Section \ref{sub:YSO-catalogs}). Furthermore, we adopt $RUWE < 1.4$ to keep those sources with well-behaved astrometrical solutions as in Method 1. We applied these cuts on the data to avoid including high uncertainty measurements, while  preserving a significant number of sources. The final sample then has 39 YSOs detected by Gaia within the G351 Environment.
    
    Fig.~\ref{fig:mosaic_kuhn} shows a concentration of YSOs in parallax for both $\varpi'$ versus $\mu$ spaces (top panels). It appears particularly clear in the parallax versus $\mu_{RA}$ panel as a compact clump of sources (top left). 
    We first identify outliers in the parallax distribution using an interquartile range criterion, to then estimate a distance from the remaining clumped sources. We consider as outliers those sources located outside the parallax range (Q1 $-f\cdot$IQR, Q3 $+f\cdot$IQR), where Q1$=0.265$, Q3$=0.703$, and IQR$=0.438$ are the quartile 1, quartile 3, and the interquartile difference Q3$-$Q1, respectively. Lastly, the factor $f=1.4$ is set slightly lower than the usual $f=1.5$ to exclude a clear outlier at $\varpi' \sim 1.3$~mas. The gray shaded area in Fig.~\ref{fig:mosaic_kuhn} (top panels) shows the extent of this parallax range, which encloses 31 clumped YSOs (refer to Table \ref{table:sources_method2} in the appendix for the astrometry used for these YSOs). Using \textit{Kalkayotl} on this sample, we obtain a distance estimation $D=1.835^{+0.235}_{-0.197} $~kpc.

Overall, our procedure is different than that adopted by \citet{kuhn21}, but it is driven by the smaller number of YSOs in our sample, together with the clear presence of (low-uncertainty) outliers and below-zero parallaxes that can heavily influence  the average result.
Unlike Method 1, the error-weighted mean parallax $\left<\varpi' \right>=$ 0.72~$\pm\,0.009$~mas of this sample suggests a much lower distance $D\simeq$~1.39~$\pm\,0.05$~kpc. On the other hand, the "geometrical" distance of \cite{Bailer-Jones_2021} suggests  $D\simeq1.80$~kpc, which is much closer to the value above and to the final value we report for G351. 
The bottom panels in Fig.~\ref{fig:mosaic_kuhn} show the proper motion (left) and spatial (right) distributions of the clumped sources and the outliers using the same symbology. Note that the right one, additionally, shows that few YSOs locate close the the G351 filament,  justifying our decision to extend the search field of YSOs towards the G351 Environment. We also note that the individual distances of the YSOs do not show any robust spatial trends or gradients between the subregions for this small sample.
    
    A last caveat is that about nine of the 31 YSOs we used to estimate a distance lie over the area of the source IRAS 17220-3609 (see the isolated contour at $b\sim 0^\circ \,20'$ in Fig.~\ref{fig:mosaic_kuhn}, bottom right panel. Also see Fig.~\ref{fig:overview}), which is unlikely to belong to the G351 molecular environment given its very different molecular line velocity \citep{leurini11b}. However, we still included them because there is no certainty about what environment they belong to.  Although it is difficult to be certain about what is the correct molecular environment of G351, we highlight that our sources still clump in parallax, which agrees with a system that is physically associated.
    
    \section{Mass and line-mass} 
    
    \subsection{Column density map}\label{method:NH2}
    
    We retrieved a total of six multi-wavelength image data (70, 160, 250, 350, 500, and 870\,$\mu$m) to reconstruct the black-body spectral energy distribution (SED) and then produce the column density N(${\rm H_2}$) map of the G351 Filament. The far-IR data from 70\,$\mu$m to 500\,$\mu$m are from the Hi-GAL survey. The 870\,$\mu$m data are from the APEX telescope large area survey of the galaxy \citep[ATLASGAL,][]{csengeri16}, which are combined with Planck data to recover the missing flux in the data processing.
    
    The data products are firstly converted into the uniform unit of ``MJy/sr'' based on the nominal beams of original images. Then we convolved the images to the same angular resolution of 45\arcsec, with a Gaussian kernel of $\sqrt{(45\arcsec)^2-\theta_\lambda^2}$ where $\theta_\lambda$ is the HPBW size of the point spread function of the radio beam. Then the convolved data from the different bands were regridded to a common pixel size of 11\parcsec5.
    
    We have used the smoothed far-IR to submillimeter image data to obtain intensity as a function of wavelength for each pixel, 
    which we model as a modified blackbody:
    \begin{equation}
        I_{\nu}= B_{\nu}(T_{\rm dust}) \left(1-e^{-\tau_\nu}\right),
    \end{equation}
    where the Planck function $B_\nu(T_{\rm dust})$ is modified by optical depth,
    \begin{equation}
        \tau_\nu = \mu_{\rm H_2} m_{\rm H} \kappa_\nu N({\rm H_2}) / R_{\rm gd}.
    \end{equation}
    Here $\mu_{\rm H_2}=2.8$ is the mean molecular weight adopted from \citet{Kauffmann2008}, $m_H$ is the mass of a hydrogen atom, $N({\rm H_2})$ is the column density of hydrogen molecule (H$_2$), and $R_{\rm gd}=100$ is the gas-to-dust ratio. The opacity $\kappa_\nu$ can be expressed as a power-law in frequency as,
    \begin{equation}
        \kappa_{\nu} = 3.33\left(\frac{\nu}{600\,\rm GHz}\right)^{\beta}\,\mathrm{cm}^{2}\,\mathrm{g}^{-1},
    \end{equation}
    where $\kappa_\nu(600\,\mathrm{GHz})=3.33$\,cm$^{2}$\,g$^{-1}$ is the dust opacity for coagulated with thin ice mantles (retrieved from column 5 of Table~1 in \citealt{Ossenkopf1994}). The dust emissivity index has been fixed to be $\beta=2.0$, in agreement with the standard value for cold dust emission \citep{Hildebrand1983}. The free parameters in the model are the dust temperature and column density. The pixelwise fitting was performed using the \texttt{least-square} method, only when the pixels have positive intensities in all the five far-IR images.  
    
    \subsection{Background subtraction and cumulative mass} \label{subsection:background_mass-convertion}
    
    We use the N(${\rm H_2}$) map to estimate the total mass of the filament and derive related metrics, such as the cumulative mass profile, the line-mass profile and the SFE. In the following we describe how these parameters are obtained. 
      
    First of all, we subtracted a uniform background contribution to the N(${\rm H_2}$) map. The subtracted level corresponds to the peak in the pixel noise distribution, which leads to a 13\% reduction of the total emission. After that, we converted N(H$_{2})$ into mass $M$ using the equation given by
    \begin{equation} \label{NH2mass}
    M =  m_H \,\mu_{H_2} \, N({\rm H_2}) \, A_{\rm pix} \, 
    \end{equation}
    where $A_{\rm pix}$ is the area of a pixel.  
    This derived mass map is the starting point of this analysis.
    
\begin{figure}
    	\includegraphics[width=\columnwidth]{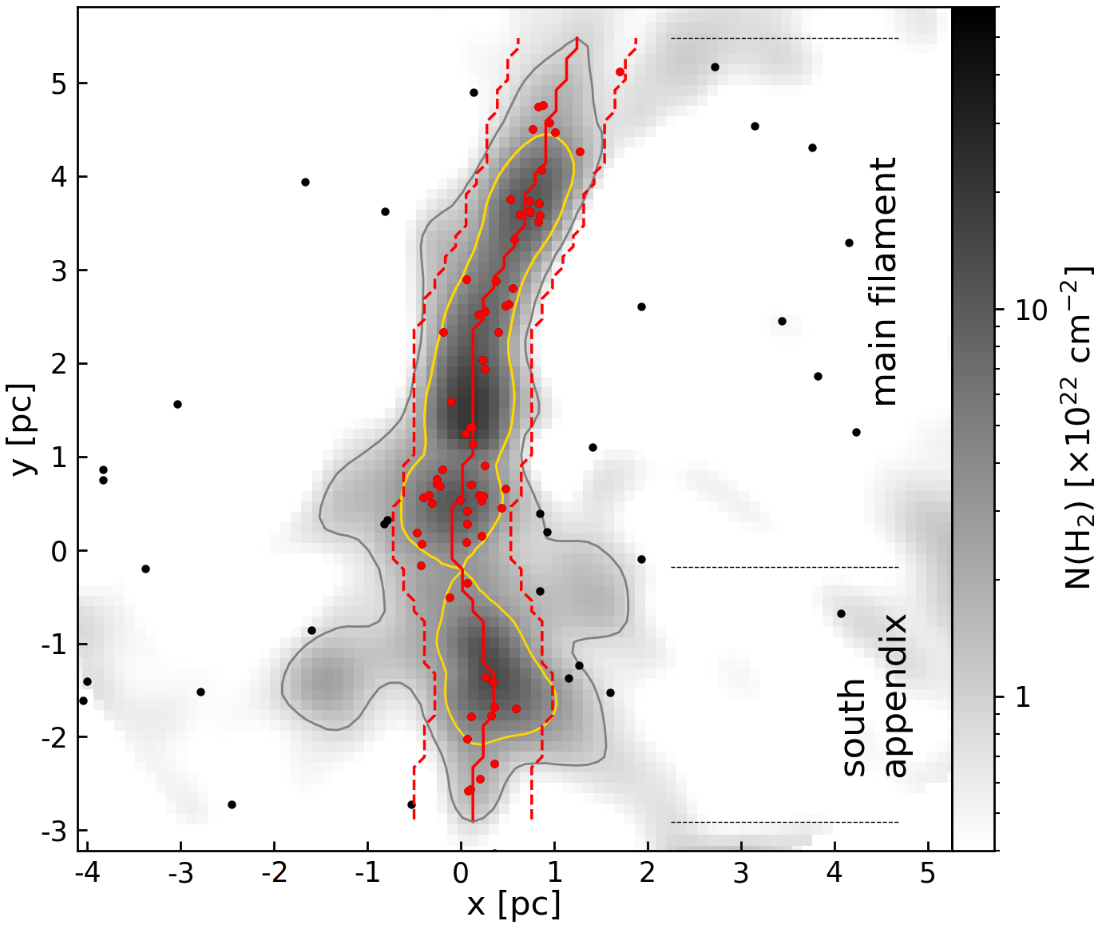}
        \caption{Rotated (see text) Herschel N(${\rm H_2}$) map of the G351 Filament. We outline the two main components of the filament (MF and SA), together with the distribution of all the 106 IR- detected YSOs. The red solid curve shows the filament's ridgeline, and the two red dashed curves are 0.63~pc offset copies of the ridgeline, representing the maximum width of the cloud (Section \ref{subsection:line-mass}). The 68 YSOs enclosed within this projected width from the ridgeline are shown with red symbols, while those outside that width are shown in black. A contour level of  9.2~$\times10^{21}$~cm$^{-2}$ (in grey) was used to define the maximum length of the filament, and the contour level 3.2$\times10^{22}$~cm$^{-2}$ (in yellow) was used to define the interface between the SA and MF.  
    	}
        \label{fig:schematic_map}
    \end{figure}

    In order to derive any mass metrics, we first rotated the map such that the longitudinal dimension of the filament is aligned as much as possible with the map pixel columns direction as shown in Fig.~\ref{fig:schematic_map}. After that, we identified the ridgeline, traced by the maximum N(H$_2$) values at each slice in the \textit{y} distribution of Fig.~\ref{fig:schematic_map}. 
    Lastly, we determined the length of the filament from a contour criteria. We used the level 9.2~$\times10^{21}$~cm$^{-2}$ (grey contour in Fig.~\ref{fig:schematic_map}) because it encloses the cloud, capturing all its features, and the pixels at both ends define the G351 filament's length as $L_{G351}=8.6$~pc. 
    
\begin{figure}
    	\includegraphics[width=\columnwidth]{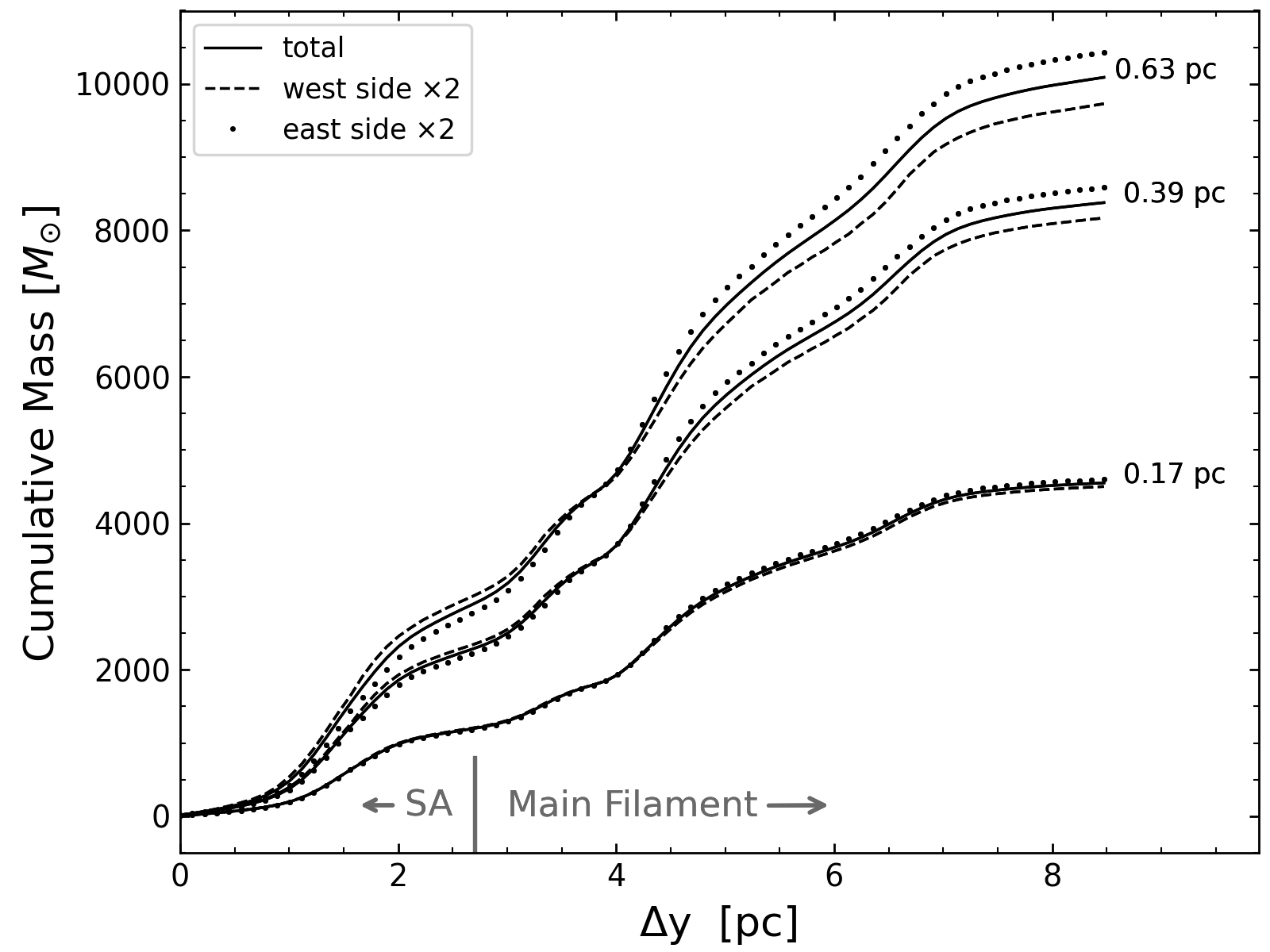}
        \caption{Enclosed cumulative mass along the G351 filament given a width~$w=$[0.17, 0.39, 0.63]~pc, counting from south to north. These three selected widths extend within the radial domain of the filament ($w~\leq~0.63$~pc), and their respective curves show similar variations, meaning that there is no significant dependency of the enclosed mass on the chosen width.}
        \label{fig:cumulative_mass}
    \end{figure}

    The next step was analyzing the longitudinal mass variations of the filament. Based on the smoothed ridgeline we re-arranged each map row horizontally to align the ridgeline pixels into the same column. Then, we calculated the cumulative mass with the varying  $y$ coordinate, summing up the mass from south to north for different widths~$w$ as shown in Fig.~\ref{fig:cumulative_mass}. This figure shows that the mass is distributed non-uniformly along the filament. We observe some jumps or curvatures due to higher mass concentrations at $\Delta y=$ 2, 3.5, 5 and 7~pc approximately, and this occurs similarly for all widths we tested. Interestingly, the location of the protocluster at $\Delta y=$ 3.5~pc is not the main jump, meaning that other portions of the filament are more concentrated.
    With this mass-declination dependency in mind, we proceed to compute the radially-projected gas line-mass profile of the cloud.
    
    \subsection{Line-mass profile, gravitational potential, and gravitational field} \label{subsection:line-mass}
    
    Following \citet{stutz16} and \citet{stutz18}, 
    the line-mass profile is the longitudinally-averaged gas distribution along a filamentary structure. Here, the cumulative mass $M$ corresponds to the total enclosed gas mass at a given projected radius from the ridgeline. This enclosed mass is then divided by that (constant) length~$L$. Formulated this way, this allows for fair comparisons between filamentary clouds of different masses and lengths.
    
\begin{figure}
    	\includegraphics[width=\columnwidth]{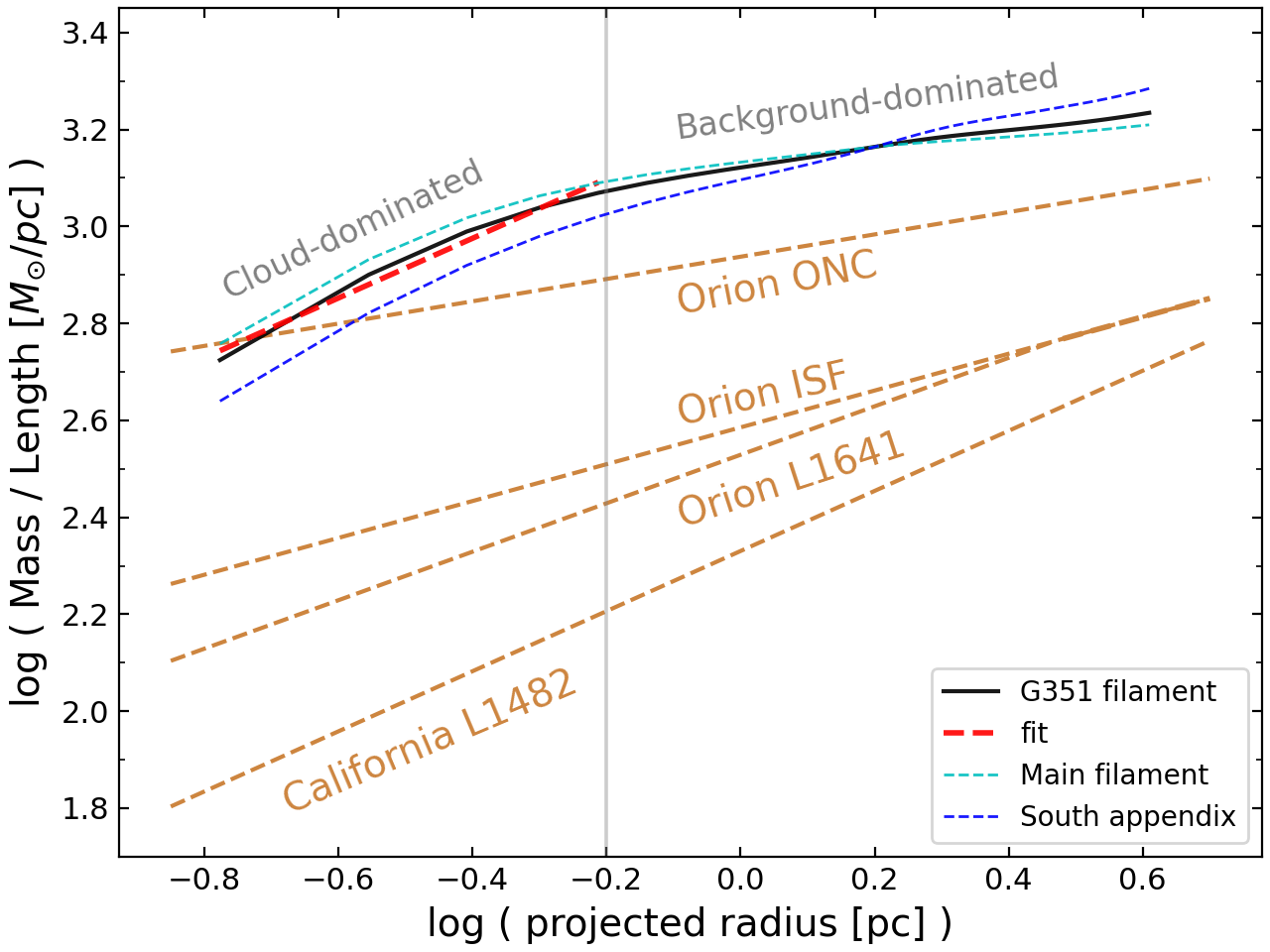}
        \caption{Enclosed gas-mass to filament-length versus projected radius from the N(H$_{2}$) ridgeline (black curve for G351). The red line is the linear fit we performed for the inner (cloud dominated) component. The profiles for both the MF and SA that make up the whole G351 filament body are included. The additional profiles of four relevant star forming regions (orange dashed lines) show that the G351 line-mass is comparatively large (Orion~A and L1482 profiles from \citealt{stutz16}, \citealt{stutz18} and \citealt{alvarez}).}
        \label{fig:linemass}
    \end{figure}

    The profile of G351 shown in Fig.~\ref{fig:linemass} (black curve) is composed of two approximately straight components (in log$-$log space), which can be approximated with a power-law following 
    \begin{equation}
    \lambda_{\rm app}(w) = \zeta \left( \frac{w}{\rm pc} \right)   ^{\gamma},
    \label{eq:linemass}
    \end{equation}
    where $\lambda_{\rm app}$ is the plane-of-the-sky (POS) projected line-mass, $w$ is the projected radius from the ridgeline, and $\zeta$ is the 1~pc normalization in units of M$_{\odot}$/pc (see Table~\ref{table:line-masses-potential}). Here, the inner curve at $r\lesssim$~0.63~pc is steeper  with slope $\gamma=0.62$, while the outer curve is flatter with slope $\gamma=0.18$. We considered that the break at $w\simeq$~0.63~pc separating both components (vertical grey line in Fig.~\ref{fig:linemass}) represents the interface at which the sky emission dominates over the measured enclosed mass. For this reason, we adopted conservatively $w=$~0.63~pc as the total radial extent (from the ridgeline) of the cloud and hence report the fitting parameters of the inner component as representing the filamentary cloud entirely (Table~\ref{table:line-masses-potential}). Our adopted total radial extent of the cloud ($2\times w$) and its length L$_{G351}=8.6$~pc enclose an area $A_{G351}=11$~pc$^2$.
    
    Given the longitudinal mass variations of the G351 filament shown in Fig.~\ref{fig:cumulative_mass}, we also estimate the individual profiles of the two main bodies that make up the whole G351 filament structure, i.e. the MF and the SA. We used the N(H$_{2}$) contour level 3.2~$\times 10^{22}$~cm$^{-2}$ (yellow contour in Fig.~\ref{fig:schematic_map}) to set the interface between the MF and the SA.
    Both profiles are slightly different to that of the total G351 filament: The MF profile is 3.5\% higher while the SA profile is 7.5 \% lower 
    (see cyan and blue profiles respectively in Fig.~\ref{fig:linemass}). 
    
    Fig.~\ref{fig:linemass} compares the G351 filament to some well-studied star forming regions in the Galaxy. Strikingly, G351 surpasses the line-mass profiles of most structures observed in Orion A and California. It is $\sim$ 3$\times$ larger than that of the ISF, and also larger than that of the ONC at $ \rm{w}\geq0.19$~pc.
    Such a high line-mass profile confirms that the G351 filament contains a concentrated reservoir of dense gas available to fuel star formation now.
    
    Line-mass {\it profiles} (or M/L profiles) are fundamental for physically characterizing molecular clouds and their constituent filaments \citep[e.g.]{stutz16,stutz18,stutz18b,alvarez}. Here, the ONC and G351 show that the two filaments have approximately similar M/L values at a specific radius, but have different profile shapes in M/L if the mass distribution is scrutinized as a function of distance from the filaments ridgeline. This information is lost when reporting a single M/L value. Overall, the above line-mass procedure has two small caveats. When projecting the original curvature of the filament towards the ridgeline, we are slightly reducing its length. Additionally, we assumed no inclination of the filament relative to the POS. These two caveats mean that our reported profile might be overestimating the actual profile of the filament, but the effect should be small. For instance, for an hypothetical inclination of 45° relative to the POS, the actual filament's length would be larger by a factor of $1/\sqrt{2}$, and then the line-mass profile would be reduced by about 30\%. Still, this is low compared to the e.g. 300\% difference relative to the ISF. 
    
    \begin{table*}
    \begin{tabular}{l c c c c c c c}  \hline
    Region  &   	$\zeta^a$		      & $\beta^b$	               & $\psi^{\,c}$ & $\xi^d$     & $\gamma^{\,\,e}$  &    Projected length & Gas mass    \\ 
       	& M$_{\odot} \rm pc^{-1}$ & M$_{\odot}\,$pc$^{-3}$ &   $(\rm{km}\,\,\rm{s}^{-1})^2$ & $(\rm{km}\,\,\rm{s}^{-1})^2 pc^{-1}$  &    &  pc & M$_{\odot}$    \\ \hline
     G351 		            & 1660 & 78.7 & 13.5 &8.38& 0.62 & 8.6  & 10200              \\  
     G351 Main Filament		& 1738    & 83.4	  &  15.1    & 8.91    & 0.59	& 5.9  & 6900               \\  
     G351 South Appendix		& 1514    &68.3     & 	10.7 &  7.28  & 0.68	& 2.7  & 2700              \\  \hline
    Orion ISF $^1$	             & 385	& 16.5 & 6.3   & 2.40 & 0.38 & 7.3  & 6200         \\
    Orion L1641 $^1$             & 338  & 16.1 & 3.50  & 1.80 & 0.50 & 23.2 & 2 $\times \, 10^{4}$  \\
    Orion ONC $^2$               & 866  & 25.9 & 27.60 & 6.40 & 0.23 & 0.5  & 1300             \\
    California L1482 $^3$        & 214  & 10.2 & 1.45  & 0.89 & 0.62 & 9.4  & 4260             \\ \hline
    \end{tabular} 
    \caption{Line-mass and star forming metrics.
    Remarks: $^{a}$for equation \ref{eq:linemass}, $^{b}$for equation \ref{eq:volumedensity}, $^{c}$for equation \ref{eq:potential}, $^{d}$for equation \ref{eq:acceleration}, $^{e}$power-law index from fitting the line-mass profile in Fig.~\ref{eq:linemass}. Comparison regions: $^{1,}$ \citet{stutz16}; $^2$\citet{stutz18};  $^3$\citet{alvarez}}
    \label{table:line-masses-potential}
    \end{table*}

    Given that the filament is highly symmetric about the ridgeline, and following \citet{stutz16} and \citet{alvarez}, we derive  from equation~\ref{eq:linemass} various profiles that are consistent with this line-mass profile and which assume cylindrical geometry.  Hence, the implied volume density $\rho_{\rm app}(r)$, gravitational potential $\Phi_{\rm app}(r)$, and the gravitational field $g_{\rm app}(r)$ profiles, respectively, are: 
    
    \begin{align}
    \rho_{\rm app}(r)= &\frac{\gamma(-\gamma/2)!}{2(-\gamma/2-1/2)!(-1/2)!} \frac{\zeta}{\rm pc^2} \left( \frac{r}{\rm pc} \right)^{\gamma-2} \nonumber \\
    =&\beta \left( \frac{r}{\rm pc} \right)^{\gamma-2}; \label{eq:volumedensity}
    \end{align}
    \begin{equation}
        \Phi_{\rm app}(r)= \psi \left(  \frac{r}{\rm pc} \right)^{\gamma}; \label{eq:potential}
    \end{equation}
    and
    \begin{equation}
        g_{\rm app}(r)= - \xi \left(  \frac{r}{\rm pc} \right)^{\gamma-1}. \label{eq:acceleration}
    \end{equation}
    We list in Table~\ref{table:line-masses-potential} the values for $\beta$, $\psi$, and $\xi$. As shown in \citet[][]{alvarez,stutz16,stutz18,stutz18b}, these profiles can be compared to other observables, such as the gas velocities in the presence of e.g., rotation and oscillations, to physically characterize the filament-averaged properties.
    
    \subsection{Total gas mass of the cloud}\label{subsection:gas-mass}
    
    With the radial extent of the cloud constrained to the inner component of the profile, we report a total gas mass $M_{gas} =$ $\sim$10$\,$200~M$_{\odot}$ ($w\leq$ 0.63~pc) for the G351 filament as a whole (see Fig.~\ref{fig:cumulative_mass}). This mass metric is about $4\times$ larger than the $\sim$2300 \Msun ($w\leq$ 0.63~pc) of the ISF (\citealt{stutz16}, see their figure 4), and larger than those of the other regions we are comparing G351 with, except L1641 which is a much larger scale structure but with a much lower line mass profile.
    The POS-projected lengths of each structure of Fig.~\ref{fig:linemass} and their respective gas masses are also listed in Table~\ref{table:line-masses-potential}. We compare our result to the literature, scaling to our assumed distance when necessary. Our gas mass agrees to within 4\% with the \cite{leurini19}, when we integrate within the same N(H$_2$) level. On the other hand, \citet{ryabukhina2021} reports a mass of 1800~M$_\odot$ (7200~M$_\odot$ scaled to our distance) using C$^{18}$O~(2-1) molecular data, so within  $\sim 30\%$ of our measurement. However, as they discuss, some negative flux features in their C$^{18}$O~(2-1) spectra might cause some underestimate.  We conclude that the agreement, despite the very different techniques, is excellent.   
    
    Overall, these results firmly establish that G351 is both more massive and compact towards its ridgeline than the ISF, and the reader may $-$ reasonably $-$ wonder at this point if the G351 star forming activity at least resembles that of the ISF (see Fig.~\ref{fig:linemass}). 
    
    \section{Discussion} \label{method: SFR-SFE} 
    
    We examine the G351 star forming activity by leveraging our catalog of YSOs and gas mass map. We start by comparing the number of YSOs per (longitudinal) parsec that G351 and the Orion A filaments are forming, taking into account incompleteness effects due to their $\sim 5 \times$ different distances and the different sensitivities of the Spitzer Survey of Orion and the Glimpse Survey. Accounting for this incompleteness will enable us  to estimate the G351 SFRs and SFEs, and will provide the means to compare the SFR and SFE per free-fall time as a function of cloud density to those values in nearby clouds.
    
    \subsection{YSO incompleteness} \label{subsec:incompleteness}
    
    The Spitzer Orion survey has provided a relatively complete survey of dusty YSOs in the Orion molecular clouds down to the hydrogen burning limit for a 1~Myr population, except in regions of bright nebulosity  \citep{megeath_2016}. As G351 is five times more distant than the Orion~A clouds ($D_{Orion\,\,A} \sim\, 390$~pc for the ISF; \citealt{kounkel17,stutz18b}), we must address the effect of incompleteness on the YSO counts measured in G351. Outside of the protocluster, the dark, high extinction regions of the cloud show a low density of sources and are devoid of bright nebulosity (see Fig.~\ref{fig:overview}, right panel). Accordingly, we do not expect source confusion to have a significant influence on the estimated YSO numbers. We therefore focus our analysis on how the lower photometric sensitivity of the Glimpse survey combined with the larger distance to G351 reduce the number of detected YSOs. We do this by using the Spitzer survey of the Orion~A cloud as a benchmark, calculating the number of Orion YSOs that would be visible at the 2~kpc distance of G351 if observed with the sensitivities of the Glimpse survey. 
    
    For this, we use the Spitzer point-source catalog of \cite{megeath12} and the  YSOs identified from that catalog published by \cite{megeath12} and \cite{megeath_2016}. For each source in the Spitzer point source catalog, we shifted the magnitude by  the difference in the distance modulus of the two clouds: 1.75~mag. We then use data from the Glimpse survey \citep{benjamin2003,churchwell2009} to find the median uncertainty at that magnitude; due to the lower integration time of the Glimpse survey these uncertainties are higher than those in the Spitzer Orion Survey.  Using this relationship between magnitude and uncertainty for all four Spitzer bands, we assigned each source in the shifted Spitzer Orion catalog the uncertainty it would have if it were in the Glimpse survey. We then rerun the YSO identification criteria from \citet{megeath12} and \citet{megeath_2016} on the IRAC data alone, ignoring criteria that use the 2MASS magnitudes or the 24~$\mu$m data. The criteria employed at this point are the same as the \cite{gutermuth09} criteria (phase-1) that we use here for Glimpse data (see Section~\ref{method:YSOidentification}). 
    
    This process provides a catalog of YSOs that would be detected by Glimpse at 2~kpc, and we notice that it keeps a fifth of the sources of the original catalog for Orion~A. We consequently apply this correction factor $k=5$ to compare line densities and to give context to the G351 SFE and SFR.
    
    \subsection{Line density of YSOs}\label{result:yso-completeness}
    
    We define the YSO line-density of a filamentary cloud as the longitudinally-averaged number of YSOs ($N_{YSO}$) that the cloud hosts per parsec, i.e.\ we divide $N_{YSO}$ by the length $L$ of the filament. Comparing G351 to the Orion~A sub-structures (ISF and L1641) through this simple metric is a revealing exercise, and it needs 3 considerations: (1) The three clouds are elongated structures that can be modeled as filaments; (2) their lengths are different; and (3) the YSO incompleteness discussed above.
    To compare line densities as fairly as possible we selected for all the clouds (G351, ISF and L1641) those YSOs within a width $w\leq 0.63$~pc from the respective ridgelines of those clouds. In addition, we considered only those YSOs that are identified by methods that use equivalent criteria. For G351 we use the phase~1 YSOs (Section \ref{method:YSOidentification}), while for Orion~A we use the YSOs that are detectable by Glimpse at 2~kpc as described in the above Section \ref{subsec:incompleteness}. We find that G351 hosts a factor of $\sim 5 \times$ fewer YSOs per parsec than the ISF, after appliying the factor $k=5$ for completeness correction (see Table~\ref{table:completeness}), revealing a lower star forming activity. Due to the possibility of incompleteness due to confusion in nebulous regions of the protocluster, we also calculate the value for the northern sub-component of the MF that excludes the protocluster (MF-north; 19 ph1-YSOs, L$=4.5$~pc) which is 4.2~YSOs~pc$^{-1}$.  We also note that in G351 MF the average YSO separation from the ridgeline is 0.177~pc. These low values for G351 suggest tension with the established SF-relations: the G351 gas is (4$\times$) more massive and (3$\times$) more compact towards its ridgeline than the ISF gas (Sections \ref{subsection:line-mass} and \ref{subsection:gas-mass}), but it is forming about 5$\times$ fewer YSOs per pc. We discuss plausible reasons for this below. 
    
    \begin{table}
    \begin{tabular}{l c c c} \hline 
     & Total N$_{YSO}$  &   Enclosed N$_{YSO}$    & Enclosed N$_{YSO}$/L  \\ \hline
    G351 & 61 & 39 & 4.5\\ 
    Orion A ISF & 195& 161 & 22\\ 
    Orion A L1641 & 114& 54 & 2.3\\ \hline
    \end{tabular}
    \caption{Number of detected YSOs in G351 compared to the completeness corrected numbers of YSOs that would be detected in Orion~A assuming a distance D$= 2$~kpc. The enclosed N$_{YSO}$ (third and fourth columns) are those within $w\leq~0.63$~pc from the ridgeline of each structure, and N$_{YSO}$/L represents the corresponding N$_{YSO}$~pc$^{-1}$. For G351, we used the phase-1 method for identifying YSOs, while for the Orion~A structures we used an equivalent method described in  \citet{megeath12} and \citet{megeath_2016} (see text). The lengths of each filament are listed in Table~\ref{table:line-masses-potential}.}
    \label{table:completeness}
    \end{table}
    
    \subsection{SFE and SFR}\label{subsection:SFR}
    
    We measure the SFE and SFR from the YSOs at $w<0.63$~pc from the G351 ridgeline. They are $N_{YSO}=68$ from all the four catalogs described in Sections \ref{sub:YSO-catalogs} and \ref{method:YSOidentification}, out of the total of 106 YSOs over the G351 Filament field (see Fig.~\ref{fig:schematic_map}).
    
    The SFE measures the fraction of the initial cloud gas mass that is converted into stars. We calculate the instantaneous SFE as \cite{myers1986} using 
$\rm{SFE}=~m_{\star}N_{YSO}/(M_{gas}+~m_{\star}N_{YSO})$, where $M_{gas}$ is the total cloud gas mass we observe today and  $m_{\star}$ is the average mass for a typical IMF ($0.5$~M$_{\odot}$). 
    Although different SFE values should be expected for different areas enclosing the cloud, we observe little variation. For the total area of the cloud ($w\leq 0.63$~pc) we find SFE~$=~0.0035$. Alternatively, for eleven N(H$_2$) contour levels spanning from 0.08 to 1$\times 10 ^{23} \rm cm^{-2}$, the SFE varies between 0.0028 and 0.0034. It can be argued that the most correct estimation for the SFE is the one that covers the total area of the cloud. For that reason, we report $\rm{SFE}=$ 0.0035 corresponding to the area within $w<0.63$~pc. 
    
    This result is an order of $10 \times$ lower than the median value SFE~=~0.038 of molecular clouds \citep{Megeath_2022}, and similarly lower to the SFE~$=0.03$ of the Orion~A cloud \citep{megeath_2016}. However, we must account for the incompleteness when comparing structures at different distances and from different surveys. In particular, we have calibrated a correction factor $k=5$ to compare G351 to Orion~A (Section \ref{subsec:incompleteness}), which we apply to $N_{YSO}$. When corrected, we find $\rm{SFE}=0.017$, which is still ($1.8 \times$) lower than that of Orion~A. Furthermore, we probe the less evolved filament portion MF-north. Within $w\leq 0.63$~pc we have 33 dusty YSOs, and a gas mass of M $= 5450$~M$_\odot$, so a low SFE$=0.003$.
    Given the high mass and high line-mass of G351, this low efficiency might be attributed to a very early evolutionary stage (see above). In that regard, magnetohydrodynamic (MHD) simulations of \citet{grudic2019} show that the instantaneous SFE continuously increases after the first sink particle forms.  Stellar feedback eventually halts the increase and the SFE remains constant at this high value. In a final note, we must bear in mind that for such a dense and massive IRDC, an above-average stellar mass $m_{\star}$ might be expected \citep{Motte2018}, hence potentially increasing our SFE estimation.
    
    On the other hand, the SFR measures the amount of solar masses that are converted into stars per unit time. We calculate it using SFR$ = m_{\star} \frac{N_{YSO}}{t_{YSO}} $. When $N_{YSO}$ is the number of protostars and disks, $t_{YSO}=2.5$~Myr because it is the expected lifetime of the disks (after which the disk disappears, \citealt{Megeath_2022}). Alternatively, when $N_{YSO}$ is the number of protostars only, $t_{YSO}=0.5$~Myr because it is the average lifetime of protostars \citep{dunham14}.
    Our YSO sample includes protostars, disks, and unclassified YSOs ($N_{YSO}=68$ for $w<0.63$~pc), hence we adopted $t_{YSO}=~2.5$~Myr to estimate SFR~=~13.6~M$_{\odot}\,$Myr$^{-1}$ for the G351 filament. This value is much lower than the 715~M$_{\odot}\,$Myr$^{-1}$ of Orion~A \citep{lada2010} and only comparable to the 16~M$_{\odot}\,$Myr$^{-1}$ found by \citet{Retes-Romero_2017} for the cloud with the lowest SFR of their sample. Finally, we probe the less evolved MF-north. Within $w\leq 0.63$~pc (33 dusty YSOs) it has SFR$=6.6$~M$_\odot$. 
    
    An essential consideration about the fixed $t_{YSO}$ we use is that it may vary depending on the environment. \citet{Bertout2007} and \citet{Galli2015} have found disk ages that depend on the mass of the parent star for two different T Tauri associations that have different SFRs. This suggests a more general dependency of the disk's lifetimes on the environment, where the average disk lifetime is longer than the 2~Myr we adopt here (for the disk-only phase) for parent star masses $>$ 0.5 M$_\odot$.  In this scenario our SFRs will vary, although still remaining significantly lower than that of Orion, which likely has a similar environment.

    \subsection{Probing local intracloud star forming relations}
    
    To confirm whether intracloud SF-relations demonstrate a unique, universal law they must be tested on different clouds. We probe the \cite{pokhrel21} relation by introducing G351 and measuring its respective $\epsilon_{\rm ff}$. Based on the three fundamental measurements of area $A$, N$_{YSO}$ and gas mass $M_{\rm gas}$ (the latter two enclosed by the given area) for different N(H$_{2})$ contour levels, we derived $\Sigma _{SFR}$, $\Sigma _{gas}$ and $t_{\rm ff}$. We follow the same procedure described in \citet{pokhrel20}, but we briefly describe it here for clarity. For a given contour, we have $\Sigma _{SFR}= \rm{SFR}/A$, $\Sigma _{gas}= M_{gas}/A$, and $t_{\rm ff}=\sqrt{3\pi/32\,G \, \rho}$, where $\rho$ is the volumetric density of the cloud assuming its mass is spherically distributed (see \citealt{hu21} for a discussion of the influence of this assumption over this relation). For the SFR they only considered protostars, hence they used $t_{YSO}=0.5$~Myr. The practical form of this \citet{pokhrel21} relation is $\rm log \,\,\Sigma_{\rm SFR}=log \left(\Sigma_{\rm gas}/ t_{\rm ff} \right) + log \,\,\epsilon_{\rm ff}$, where $\epsilon_{\rm ff}$ is the fraction of gas mass converted to stars per free-fall time. 
    
    For G351 we did not limit our sample to only protostars because not all the catalogs we retrieved YSOs from provide a classification. Instead, we used our entire sample of YSOs (Sections \ref{sub:YSO-catalogs} and \ref{method:YSOidentification}), which includes protostars and disks, and accordingly used $t_{YSO}=2.5$~Myr (see discussion about this fixed value in Section \ref{subsection:SFR}). For all the clouds we selected similar  contour levels that range from 0.82 to 4.82$\times 10^{22}$~cm$^{-2}$ in steps of 4$\times10 ^{21}$~cm$^{-2}$, and which were selected within the resolution limitations of our N(H$_{2})$ map. 
    We also replicate the Orion~A cloud and its two main sub structures (ISF and L1641), using the column density N(H$_{2})$ of \cite{stutz16} for the mass measurements, together with the declination limits they defined for each structure. For the SFR calculations, we used the Spitzer Orion Survey YSOs and the protostars of \citet{megeath12,megeath_2016} because it is the most complete catalog over that area and for consistency with all analysis throughout this section. We note that when using (an slightly updated version of) the protostars catalog that \citet{pokhrel21} used (Pokhrel 2023, priv.\ comm.), we very closely reproduce their Orion~A profile, obtaining $\epsilon_{\rm ff}= 0.0086$ (14\% lower than their $\epsilon_{\rm ff}= 0.001$).
    For the ISF and L1641 individually we estimated $\epsilon_{\rm ff}= 0.0064$ and $\epsilon_{\rm ff}= 0.008$, respectively.
    
\begin{figure}
    	\includegraphics[width=\columnwidth]{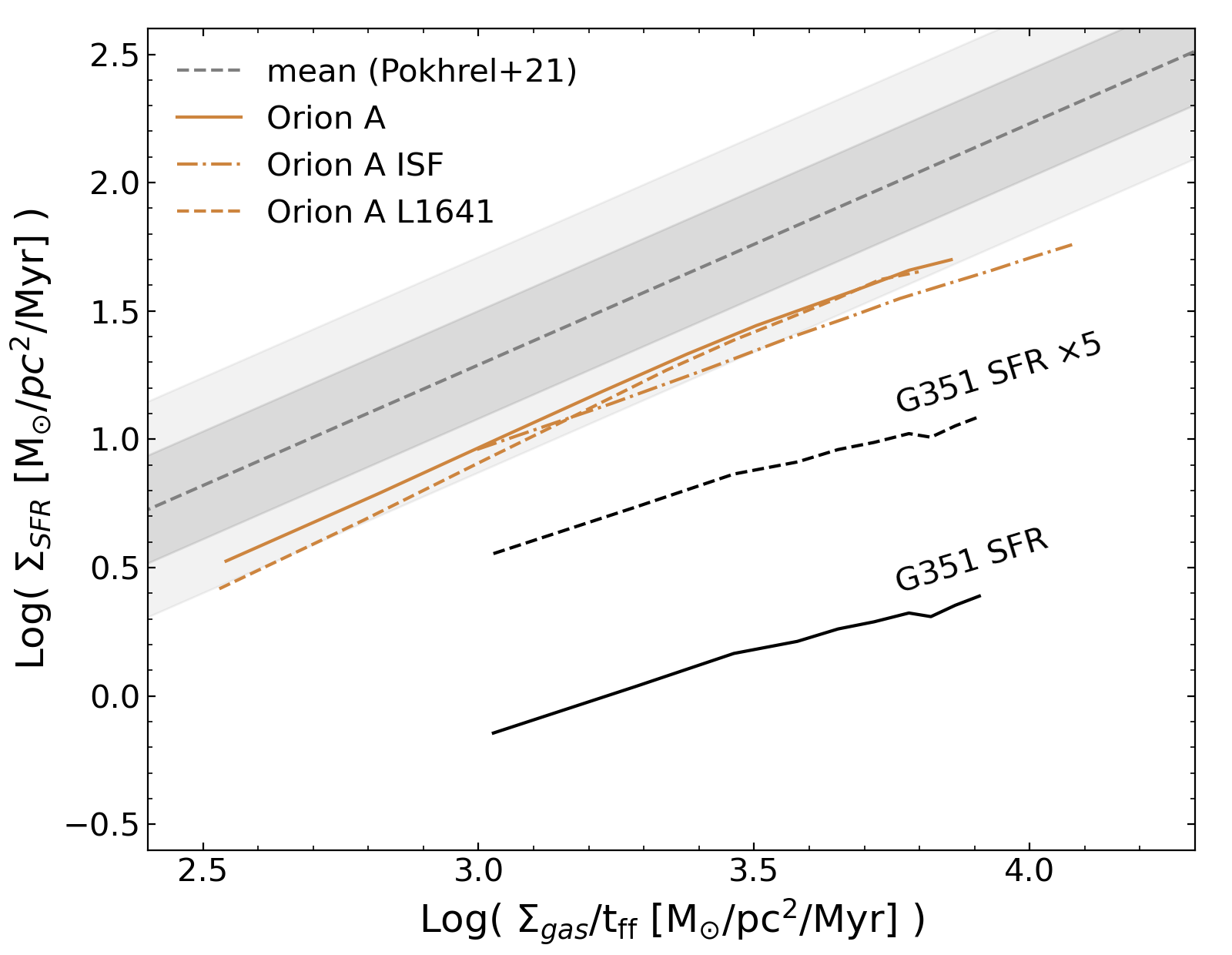}
        \caption{Grey-shaded area: SF-relation from \citet{pokhrel21}, showing a linear-law where all clouds have a similar efficiency per free-fall time $\epsilon_{\rm ff}$. Orange curves show Orion~A and its sub-regions separately. When G351 is added to this diagram, the measured SFR implies a lower value of $\epsilon_{\rm ff}$ than that of nearby clouds, even with our completeness correction $k=5$.}
        \label{fig:pokhrel}
    \end{figure}

    In Fig.~\ref{fig:pokhrel} the 12 clouds of \cite{pokhrel21} are represented by their mean and dispersion. We also include here an scaled G351 profile that accounts for the YSO incompleteness relative to the Orion~A clouds, where we increased $N_{\rm YSO}$ in the SFR term by our factor $k=5$ (Section~\ref{result:yso-completeness}). We measured a "raw" $\epsilon_{\rm ff}=0.00042$ for the G351 filament. Moreover, after completeness correction, $\epsilon_{\rm ff}=0.002$.  This value is lower than both the mean intracloud relation by a factor of 12 and that of the lowest cloud (Orion~A) in the \cite{pokhrel21} sample by a factor of $\sim\,4.7$. 
    To obtain the latter values we averaged over the entire filament structure, but it is also informative to probe the less evolved north component (i.e. MF-north, Section \ref{result:yso-completeness}) that contains the most quiescent clumps \citep{leurini19}. This portion of the MF shows a lower $\epsilon_{\rm ff}$ ($=0.0003$) relative to G351 as a whole by a factor of $\sim\,$0.7. This highlights a significant difference in evolutionary stages throughout different environments within the G351 filament.   
    
    Our analysis demonstrates that the global star-forming activity in the G351 filament is low. To explain this we can argue a few potential reasons. First, the cloud may be less than 2.5~Myr in age, which would also result in an underestimate of the SFR, but would predict an elevated number of protostars compared to the more evolved pre-main sequence stars (e.g. \citealt{stutz15}, see models in \citealt{Megeath_2022}).  For the youngest clump in the MF, \citet{Giannetti19} measured an age $\leq 10^{5}$~yr from mm-wave emission of chemical tracers. Second, the lower $\epsilon_{\rm ff}$ might indicate that it varies with environment in our Galaxy . Third,  mechanisms like B-fields or turbulence might be supporting the cloud against collapse. However, this last possibility, that of turbulence, would imply high line-widths to support the mass, line-mass, and volume densities calculated above. In this regard, the observation from \citet{Giannetti19} that at least "Clump 7" is sub-virial based on APEX observations of o-H$_2$D$^+\,(1_{1,0} - 1_{1,1})$ line widths may further call into question large contributions of turbulent support at least on the scales of clumps.  Moreover, as discussed above, the \citet{ryabukhina2021} M/L value (scaled to our distance) would imply a sub-virial (relative to radial collapse) filament, as opposed to the equilibrium values these authors tentatively infer.  Indeed, as \citet{ryabukhina2021} point out (along with a careful discussion of possible uncertainties), despite the measurement of radial near-equilibrium, the G351 filament is in fact observed to be fragmenting and collapsing right now. Lastly and regarding the $\epsilon_{\rm ff}$ measurement in particular, the assumption that the cloud is spherical might have played a role as the gas distribution is elongated and more likely characterized by a cylindrical geometry (see above).  
        
    \section{Conclusions}
    
    We have characterized the global star forming properties of the G351 IRDC by systematically comparing to the Orion~A cloud, and particularly the ISF, the best nearby reference for cluster formation in filaments. Based on the  measurements of distance, gas mass, and YSO counts, we establish a solid comparative frame that allows us to (a) constrain its current star forming activity and (b) test intracloud SF- relations.
    
    We first constrain the G351 distance following two Gaia DR3 based approaches. In method 1 we find sources clumping in proper motion space for only one narrow parallax range: $\varpi'~\in~[0.45,0.50]$~mas. From that clump we estimate $D=~2.109^{+0.267}_{-0.220}$~kpc. In method 2 we consider YSOs that are Gaia-detected, and to maintain a relevant number of sources we probe the entire environment of G351. This field contains bright clouds that are plausibly associated to the filament, and in agreement with this association hypothesis we find that the YSOs clump in parallax. From that clump we estimate $D=1.835^{+0.235}_{-0.197} $~kpc. As the two methods retrieve similar results we take their average and report the first Gaia-based distance to G351 $D=2 \pm 0.14$~kpc, resolving at the same time the long-standing controversy about the G351 distance.
    
    With this distance, we are then able to measure gas masses. We calculate mass from our column density N(H$_2$) map created from Herschel/APEX imaging. We first identify the ridgeline of the filament, to then define its total length $L=8.6$~pc, and conservatively define a total radial extent $w=0.63$~pc, where $w$ is the POS projected radius from the ridgeline of the filament. Within these dimensions there is an area of 11~pc$^2$ that encloses a total gas mass of 10$\,$200~M$_{\odot}$.
    Beyond an absolute mass measurement, we focus on the $-$ more comprehensive $-$ distribution of the gas mass. We find that the G351 line-mass follows a simple power-law $\lambda = 1660  (w/\rm pc )^{0.62} \,\, M_{\odot}\,\rm{pc}^{-1}$, higher than that of all Orion~A structures and California L1482 at all widths $w$ where the profiles can be compared, except in the very center of the ONC (M42) at $w\leq 0.19$~pc. In particular, our $\lambda$ is a factor of 3 greater than that of the ISF. In addition, we find that the power-law index $\gamma=0.62$ of our $\lambda$ profile is larger than all the Orion structures but similar  to that of the California L1482 filament. All the above confirms that the G351 filament contains a concentrated reservoir of dense gas that 
    has the potential to fuel active star formation.
    
    In contrast, we measure a low star formation activity based on the YSOs found in the filament area. Our initial analysis of line YSO densities reveal that G351 is forming 5$\times$ fewer YSOs per parsec than the ISF, and when including mass, we find that the SFE is 1.8 times lower.  For both results we consider the YSO incompleteness correction between G351 and the Orion~A cloud (a factor $k=5$). Additionally, by measuring SFRs (considering incompleteness) and gas masses per free-fall time we test star forming relations for local clouds. Here we measured a "raw" $\epsilon_{\rm ff}= 0.0004$ for the filament, which becomes $\epsilon_{\rm ff}= 0.002$ after completeness correction. We show that the low star formation activity of this IRDC is significantly below the SF-relation of \citet{pokhrel21}, with  $>$2$\sigma$ discrepancy from mean local 12-cloud relation.  Specifically, we measure a 1.1~dex lower $\epsilon_{\rm ff}$ than the mean local intracloud relation reported by \citet{pokhrel21}, and a factor of 4.7 times lower than the least efficient cloud in that analysis, the Orion~A cloud, which we also reproduce independently. This suggest that intracloud SF-relations do not capture the variations of properties of IRDCs relative to nearby star forming clouds.
    
    Given its relatively large and concentrated reservoir of gas and hence its strikingly low star formation activity demonstrated here, this cloud begs for an explanation of what physical conditions produce such an inefficiency. One potential explanation is that G351 could be very young. However, it has already initiated cluster formation, so this hypothesis must be regarded with some caution unless one assumes rapid cluster formation or different ages of emergent start structures in the same maternal filament.  Alternatively, it is possible that environmental mechanisms are supporting it against collapse, such as magnetic fields or turbulence. Regarding the latter, recent measurements of dense tracers in portions of the filament suggest it to be sub-virial \citep[][scaling to our distance]{Giannetti19,ryabukhina2021}, hence turbulence appears a less likely candidate for supporting the cloud. 
    
    The low age hypothesis can be tested by further studies that identify the most embedded YSOs along the filament and in the active protocluster (e.g., using the James Webb Space Telescope), by measuring the ratio of the number of pre-main sequence stars with disks (Class II) to protostars (Class I) \citep{Gutermuth_2011,stutz15,Megeath_2022}. 
    We can further leverage high resolution ALMA detections of cores in the protocluster \citep{motte22} and in the filament as a whole by determining the fraction of cores with protostars \citep{nony23}. This would establish their evolutionary stage, and in turn estimate their evolutionary timescales (albeit with small number statistics), calibrated to our current understanding of YSO and protostar lifetimes.  
    Lastly, the magnetic field hypothesis should  be tested with polarimetry. In particular,  linear dust polarization is becoming increasingly observationally accessible (via e.g., ALMA, APEX, LCT, or CCAT-prime, to mention a few southern observatories that either have or will have polarization capabilities), and provides constraints on the geometry of the field, for example, that may depend on environment (main protocluster versus the rest of the filament).
    
    \section*{Acknowledgements}
    The authors are very grateful to T. Csengeri, A. Ginsburg, and P. Sanhueza for their constructive comments.    
    SR gratefully acknowledges the funding and support of ANID-Subdirección de Capital Humano Magíster/Nacional/2021-22211000. AS gratefully acknowledges support by the Fondecyt Regular (project code 1220610), and ANID BASAL project FB210003. S.T.M. gratefully acknowledges funding support from the NASA/ADAP grants 80NSSC18K1564 and 80NSSC20K0454, as well as NSF AST grant 2107827. During part of this work, S.T.M. was a Fulbright Scholar at the Universidad de Concepción, Chile. FWX acknowledges the National Science Foundation of China (12033005). H.-L. Liu is supported by National Natural Science Foundation of China (NSFC) through the grant No.12103045, and by Yunnan Fundamental Research Project (grant No.\,202301AT070118).  
    R.A. gratefully acknowledges support from ANID Beca Doctorado Nacional 21200897.
    This work has made use of data from the European Space Agency (ESA) mission Gaia (\url{https://www.cosmos.esa.int/gaia}), processed by the {\it Gaia} Data Processing and Analysis Consortium (DPAC,
    \url{https://www.cosmos.esa.int/web/gaia/dpac/consortium}). Funding for the DPAC has been provided by national institutions, in particular the institutions participating in the {\it Gaia} Multilateral Agreement.
    
    
    \section*{Data Availability}
    Data directly related to this publication are available upon reasonable request to the corresponding author.
    
    
    \bibliographystyle{mnras}
    \bibliography{ref} 
    
    
    
    
    \appendix
\section{Sources used in Method 1 and Method 2}
\begin{table*}
    \centering
    \begin{tabular}{cccccc}  \hline
        source\_id&ra&dec&pmra&pmdec&parallax$^1$   \\ 
        & [deg] & [deg] & [mas$\,$yr$^{-1}$] & [mas$\,$yr$^{-1}$] & [mas]   \\  \hline
5974562944852710016& 261.67278&-36.13344 &-1.438 &-2.205&0.491 \\
5974563052229952640& 261.68242 & -36.11678 &-1.238 &-2.533&0.461\\
5974566002869457792& 261.67245&-36.11122 &-1.259 &-2.453 &0.490\\
5974560131652197632& 261.59294 &-36.14386 &-0.992 &-3.344 &0.471\\
5974556867473293312& 261.61314&-36.18537&-0.847&-2.440&0.441\\
5974566518265600640& 261.68802 & -36.06515 &-1.588 &-3.043&0.454\\
5974563528977056384&261.74423&-36.08853&-1.125&-2.881 &0.419\\
5974563567626287616&261.74270&-36.07842&-1.508 &-2.762 &0.468\\
5974556214638241408&261.63190 &-36.21996 &-1.785 &-2.755 &0.436\\
5974560917628712192&261.55679&-36.13589 &-0.731 &-2.485 &0.450\\
5974560608390516608&261.53823 & -36.12648&-1.676 &-2.600 &0.489\\
5974561192506102144&261.53947 &-36.10154 &-1.876 &-2.620&0.465\\
5974514673714527872&261.77084 &-36.20683 &-1.109 &-3.096 &0.471\\
5974567656435178112&261.67853 &-36.02312 &-1.452 &-3.280&0.476\\
5974556111562786816&261.62245&-36.25135&-1.180 &-3.564&0.490\\
5974565148170946816&261.75793 &-36.03345&-1.223&-2.761 &0.454\\
5974514463269040128&261.75760 &-36.24571 &-0.974 &-2.510 &0.437\\
5974556313430219776&261.57277 &-36.24542 &-0.063 &-2.470 &0.394\\
5974567823935744640&261.64413 &-36.00384&-1.401 &-3.099 &0.488\\
5974568373691971968&261.71957&-36.00846&-1.335 &-2.766 &0.488\\
5974569163965546880&261.69948&-35.99930 &-1.513 &-3.429 &0.477\\
5974568377989684096&261.72216&-36.00270 &-1.073 &-2.913 &0.457 \\ \hline
    \end{tabular}
    \caption{Gaia DR3 parameters used in Method 1, for all the 22 sources used in the distance estimation. Remarks:     $^1$ Corrected parallax $\varpi'$.}
    \label{table:sources_method1}
\end{table*}


    \begin{table*}
    \centering
    \begin{tabular}{cccccc}  \hline
      source\_id&ra&dec&pmra&pmdec&parallax$^1$ \\ 
      & [deg] & [deg] & [mas$\,$yr$^{-1}$] & [mas$\,$yr$^{-1}$] & [mas]  \\  \hline
5974506972837549568&261.72842 &-36.35586 &-4.189&-6.762 &0.602 \\
5974508622105811456&261.72253&-36.25625&0.827 &-9.738 &0.871\\
5974514639358565376&261.78857&-36.20896 &-3.157&-3.517 &0.642\\
5974529169228342144&261.53774 &-36.42848&-0.224&-2.600 &0.489\\
5974530509270521216&261.45756 &-36.36795 &-1.069 &-3.602 &-0.085\\
5974530646697360000&261.47316&-36.33342&1.128&2.391 &0.410\\
5974533601642040192&261.46084&-36.33869&-0.674 &-1.980 &0.186\\
5974533915171434880&261.41665&-36.30318&-1.006&0.954 &0.547\\
5974548101453367808&261.27705&-36.18545&-0.740 &-1.596 &0.423\\
5974551060681037056&261.26908 &-36.17087&0.228&-2.764 &0.816\\
5974551193829822208&261.27840&-36.14585&-0.356 &-0.494 &0.206\\
5974552808728470016&261.58265&-36.38151&-1.310 &-5.584 &0.552\\
5974553431502710272&261.61030&-36.33359 &0.623 &-0.267&0.530\\
5974554874612191872&261.65887 &-36.32143&-3.460 &-2.870 &0.821\\
5974555007765299584&261.67202 &-36.30336 &2.211 &1.060 &0.456\\
5974555351349424640&261.63329 &-36.26256&-6.254&-3.979 &0.418\\
5974556519588723328&261.65666 &-36.22673 &-5.276 &-4.169 &-0.153\\
5974556936196518272&261.65251 &-36.17377&-1.570&-4.371 &0.652\\
5974559959853752960&261.62330 &-36.15631&-2.354 &-12.412 &0.908\\
5974560750127755392&261.50222&-36.13253 &-5.223&-17.718 &0.054\\
5974567342907875200&261.58554&-36.03254&-6.023&-2.820 &0.532\\
5974570508292884352&261.41751 &-36.18073&-0.748&-2.725 &0.503\\
5974570611372116864&261.35724&-36.19348&-1.068&-1.514 &0.404\\
5974571917042192512&261.39029&-36.13617&-1.608&-4.176 &-0.069\\
5974572226276449408&261.37033&-36.10281 &3.866 &-0.147 &0.401\\
5974573699450931968&261.44567&-36.06172&-9.196 &-14.523 &0.504\\
5974574214846730752&261.48483&-36.01846 &1.302 &-2.070 &0.754\\
5974560402232025856&261.53440&-36.16568&-0.2409 &-0.954 &0.325\\
5974570542652628992&261.36880&-36.20487&0.230&-0.999 &0.638\\
5974508037989573632&261.71023&-36.30577&-0.962&-0.731 &-0.025\\
5974557378574537728&261.45825&-36.28780&0.939&-0.279 &0.517\\ \hline
    \end{tabular}
    \caption{Gaia DR3 parameters of the 31 YSOs used in Method 2. In this case, proper motions are included for completeness but are not used to estimate a distance. Remarks: $^1$ Corrected parallax $\varpi'$.}
    \label{table:sources_method2}
\end{table*}

    
    

    \label{lastpage}
    \end{document}